\documentclass[prd,preprint,superscriptaddress,preprintnumbers,eqsecnum,showpacs,nofootinbib,nobibnotes,noeprint]{revtex4-1}
\usepackage{amsfonts,amsmath,mathbbol,amssymb,bm,natbib}
\usepackage[table,dvipsnames]{xcolor}
\usepackage{graphicx} 
\usepackage{color}
\usepackage{stackengine}
\usepackage{scalerel}
\usepackage{mathtools}
\usepackage[linkcolor=blue,citecolor=blue,urlcolor=blue,colorlinks=true]{hyperref}

\definecolor{refkey}{rgb}{0.39,0.58,1}
\definecolor{labeled}{rgb}{1,0,0}

\usepackage{multirow}
\usepackage{mathrsfs}
\usepackage{array}

\usepackage{physics}

\def\ie{{\it i.e.}, }
\newcommand{\be}{\begin{equation}}
\newcommand{\bea}{\begin{eqnarray}}
\newcommand{\ee}{\end{equation}}
\newcommand{\eea}{\end{eqnarray}}
\newcommand{\fatg}{{\rm{I}}\!\Gamma} 

\def\s#1{{\scriptscriptstyle #1}}
\def\srm#1{{\rm{\scriptscriptstyle #1}}}

\newcommand{\Dr}{\mathcal Z}     

\def\1eq#1{Eq.~(\ref{#1})}

\def\2eqs#1#2{Eqs.~(\ref{#1}) and~(\ref{#2})}
\def\3eqs#1#2#3{Eqs.~(\ref{#1}),~(\ref{#2}) and~(\ref{#3})}

\def\eg{{\it e.g.}, }
\newcommand{\ols}[1]{\mskip.5\thinmuskip\overline{\mskip-.5\thinmuskip {#1} \mskip-.5\thinmuskip}\mskip.5\thinmuskip} 

\newcommand{\Ls}{ \mathit{L}_{{sg}}}                         
\newcommand{\TLs}{ \mathit{{L}^{\scriptscriptstyle \rm T}}_{{\!\!\!\!sg}}}               

\begin{document}

\title{Ghost dynamics in the soft gluon limit}

\author{A.~C.~Aguilar}
\affiliation{\mbox{University of Campinas - UNICAMP, Institute of Physics ``Gleb Wataghin'',} 
13083-859 Campinas, S\~{a}o Paulo, Brazil.}

\author{C.~O. Ambr\'osio}
\affiliation{\mbox{University of Campinas - UNICAMP, Institute of Physics ``Gleb Wataghin'',} 
13083-859 Campinas, S\~{a}o Paulo, Brazil.}

\author{F.~De Soto}
\affiliation{\mbox{Dpto. Sistemas F\'isicos, Qu\'imicos y Naturales, Univ. Pablo de Olavide, 41013 Sevilla, Spain.}}

\author{M.~N. Ferreira}
\affiliation{\mbox{University of Campinas - UNICAMP, Institute of Physics ``Gleb Wataghin'',} 13083-859 Campinas, S\~{a}o Paulo, Brazil.}

\author{B.~M. Oliveira}
\affiliation{\mbox{University of Campinas - UNICAMP, Institute of Physics ``Gleb Wataghin'',} 13083-859 Campinas, S\~{a}o Paulo, Brazil.}

\author{J.~Papavassiliou}
\affiliation{\mbox{Department of Theoretical Physics and IFIC,} \\ University of Valencia and CSIC, E-46100, Valencia, Spain.}

\author{J.~Rodr\'{\i}guez-Quintero}
\affiliation{\mbox{Department of Integrated Sciences, University of Huelva, E-21071 Huelva, Spain.}}

\pacs{
12.38.Aw,  
12.38.Lg, 
14.70.Dj 
}

\begin{abstract}

We present a detailed study of the dynamics associated with the ghost sector of quenched QCD in the Landau gauge,
where the relevant dynamical equations are supplemented 
with key inputs originating from large-volume lattice simulations.
In particular, we solve the coupled system of Schwinger-Dyson equations that governs the evolution of the
ghost dressing function and the ghost-gluon vertex, using as input for the gluon propagator
lattice data that have been cured from volume and discretization artifacts. 
In addition, we explore the 
soft gluon limit of the same system, employing recent lattice data 
for the three-gluon vertex that enters in one of the diagrams defining the
Schwinger-Dyson equation of the ghost-gluon vertex.
The results obtained from the numerical treatment of these equations
are in excellent agreement with lattice data for the ghost dressing function, once the latter have 
undergone the appropriate scale-setting and artifact elimination refinements. 
Moreover, the coincidence observed between the ghost-gluon vertex 
in general kinematics and in the soft gluon limit reveals an outstanding consistency 
of physical concepts and computational schemes.
  
\end{abstract}

\maketitle

\section{\label{sec:intro}Introduction}

In the ongoing quest for unraveling the nonperturbative structure of QCD, considerable effort has been 
dedicated to the study of Green's (correlation) functions by means of both continuous methods~\cite{Roberts:1994dr,Alkofer:2000wg,Maris:2003vk,Pawlowski:2003hq,Pawlowski:2005xe,Fischer:2006ub,Aguilar:2006gr, Roberts:2007ji,Aguilar:2008xm, Boucaud:2008ky,Fischer:2008uz,Binosi:2009qm,Tissier:2010ts,Campagnari:2010wc,Pennington:2011xs,Aguilar:2011xe,Vandersickel:2012tz,Serreau:2012cg,Fister:2013bh,Cloet:2013jya,Binosi:2014aea,Kondo:2014sta,Aguilar:2015bud,Binosi:2016rxz,Binosi:2016nme,Corell:2018yil,Cyrol:2017ewj,Gao:2017uox,Huber:2018ned,Pelaez:2021tpq}  and large-volume lattice simulations~\cite{Sternbeck:2005tk,Ilgenfritz:2006he,Cucchieri:2007md,Cucchieri:2007rg,Bogolubsky:2007ud,Bowman:2007du,Cucchieri:2008fc,Cucchieri:2009zt,Bogolubsky:2009dc,Oliveira:2009eh,Oliveira:2010xc,Maas:2011se,Boucaud:2011ug,Oliveira:2012eh,Ayala:2012pb,Bicudo:2015rma}.
In this pursuit, the detailed scrutiny of the ghost sector of the theory    
is particularly important, both because of its direct connection with specific scenarios of color confinement  \cite{Kugo:1979gm,Nakanishi:1990qm},
but also due to its impact on the nonperturbative    
behavior of other key Green's functions, such as the gluon propagator and the three-gluon vertex~\cite{Cucchieri:2006tf,Cucchieri:2008qm,Alkofer:2008jy,
Huber:2012zj,Aguilar:2013vaa,Pelaez:2013cpa,Blum:2014gna,Eichmann:2014xya,Williams:2015cvx,Blum:2015lsa,Cyrol:2016tym,Duarte:2016ieu,Athenodorou:2016oyh,Boucaud:2017obn,Aguilar:2019jsj,Aguilar:2019uob,Aguilar:2019kxz}.
In particular, the nonperturbative masslessness of the ghost is responsible for the vanishing   
of the gluon spectral density at the origin~\cite{Cyrol:2018xeq,Haas:2013hpa,Fischer:2020xnb,Horak:2021pfr},  and for the infrared suppression of the three-gluon vertex~\cite{Huber:2012zj,Aguilar:2013vaa,Athenodorou:2016oyh,Boucaud:2017obn,Aguilar:2019jsj}. 
In that sense, the ghost dynamics leave their imprint on a variety of fundamental phenomena, 
such as chiral symmetry breaking and the generation of quark constituent masses\mbox{~\cite{Aguilar:2018epe,Gao:2021wun,Mitter:2014wpa,Aguilar:2010cn,Fischer:2003rp,Roberts:1994dr}},
the emergence of a mass gap in the gauge sector of the theory~\cite{Cornwall:1981zr,Aguilar:2008xm,Aguilar:2019kxz,Aguilar:2020uqw},  and the dynamical formation of hadronic bound states~\cite{Maris:2003vk,Maris:1997tm,Cloet:2013jya,Eichmann:2009qa,Eichmann:2016yit}  and glueballs~\cite{Meyers:2012ka,Fukamachi:2016wxf,Souza:2019ylx,Huber:2020ngt}.

In the framework of the Schwinger-Dyson equations (SDEs),
the momentum evolution of the ghost dressing function is governed by a
relatively simple integral equation, whose main ingredients are the
gluon propagator and the fully-dressed ghost-gluon vertex.
If one treats the 
gluon propagator as external input obtained from lattice simulations (see \eg~\cite{Aguilar:2013xqa}),
then the main technical challenge of this approach is the determination of the ghost-gluon vertex.
In the Landau gauge, the ghost-gluon vertex is rather special, because, by virtue of Taylor's theorem, 
its renormalization constant is finite~\cite{Taylor:1971ff}.  Of the two possible tensorial structures allowed by Lorentz invariance, only that corresponding to the classical (tree-level) tensor survives in the
calculations. The form factor associated with it will be denoted by $B_1(r,p,q)$, where
$r$, $p$, and $q$ are the momenta of the antighost, ghost, and gluon, respectively.

The most complicated aspect of the SDE that determines $B_1(r,p,q)$
is that, in addition to  $B_1(r,p,q)$ itself, the resulting integral equation,
derived in the so-called  ``one-loop dressed'' approximation, 
depends also on the fully-dressed three-gluon vertex.
This latter vertex has a rich tensorial structure~\cite{Ball:1980ax}, 
and a complicated description at the level of the SDEs\mbox{~\cite{Schleifenbaum:2004id,Huber:2012kd,Aguilar:2013xqa,Huber:2012zj,Blum:2014gna,Eichmann:2014xya,Williams:2015cvx, Binosi:2016wcx,Hawes:1998cw,Chang:2009zb,Qin:2011dd};} 
therefore, it is often approximated by resorting to gauge-technique constructions~\cite{Salam:1963sa,Salam:1964zk,Delbourgo:1977jc,Delbourgo:1977hq}, 
based on the Slavnov-Taylor identities (STIs) that it \mbox{satisfies}.

The comprehensive treatment of the relevant SDEs presented in~\cite{Aguilar:2018csq}
gives rise to a $B_1(r,p,q)$  with a mild momentum-dependence and a modest
deviation from its tree-level value (see also~\cite{Aguilar:2013xqa}), and a ghost dressing function, $F(q^2)$, that is in good
(but not perfect, see, \eg right panel on \mbox{Fig.~16} of~\cite{Aguilar:2018csq}) agreement with the lattice data~\cite{Bogolubsky:2009dc}.

In the present work,  we take a fresh look at the system of coupled SDEs that determines 
the ghost dynamics, taking advantage of two recent advances in the area of lattice QCD~\mbox{\cite{Aguilar:2021lke,Boucaud:2017ksi, Boucaud:2018xup}}.
First, the simulation of the three-gluon vertex in the ``soft gluon limit'' \mbox{($q \to 0$)}~\cite{Aguilar:2021lke}  
furnishes accurate data for a special form factor, denoted by $\Ls(r^2)$, which
constitutes a central ingredient of the SDE for $B_1(r,p,q)$,
when computed in the same kinematic limit, namely $B_1(r,-r,0)$. 
Second, the lattice two-point functions employed in our study  
have been cured from volume and discretization artifacts,
once the scale-setting and continuum-limit extrapolation 
put forth in~\cite{Boucaud:2017ksi,Boucaud:2018xup} have been implemented.

The way the aforementioned elements are incorporated into the present analysis is as follows.
The starting point is the computation of $F(q^2)$ and $B_1(r,p,q)$ from the
coupled system of SDEs they satisfy. In the SDE for $B_1(r,p,q)$,  an approximate form of 
the three-gluon vertex is employed: only the tree-level tensorial structures are retained, and the 
associated form factors are taken from the STI-based derivation of~\cite{Aguilar:2019jsj}. 
In addition, 
the gluon propagator of~\cite{Boucaud:2018xup} combined with that of \cite{Bogolubsky:2009dc}, subjected to the refinements mentioned above, is used in the SDEs as external input.
The solution of the system yields a $F(q^2)$ which is in outstanding agreement with the ghost dressing
function of~\cite{Boucaud:2018xup}. The corresponding
solution for $B_1(r,p,q)$, in general kinematics,  displays the salient features known from previous studies~\cite{Schleifenbaum:2004id,Huber:2012kd,Aguilar:2013xqa,Cyrol:2016tym,Mintz:2017qri,Aguilar:2018csq,Huber:2018ned,Aguilar:2019jsj,Barrios:2020ubx}.
In fact, one may extract from it various kinematic limits as special cases, 
and, in particular, the two-dimensional ``slice'' that corresponds to the
soft gluon limit, thus obtaining $B_1(r,-r,0)$.  

The next step is to implement the soft gluon limit \mbox{($q \to 0$)} {\it directly} at the level of the
SDE for $B_1(r,p,q)$, which is thus converted to a dynamical equation for $B_1(r,-r,0)$.
By virtue of this operation, the three-gluon vertex nested in one of the defining Feynman diagrams
is projected naturally to its soft gluon limit, thus allowing
us to replace it precisely by the function $\Ls(r^2)$
obtained from the lattice analysis of~\cite{Aguilar:2021lke},
without having to resort to any Ans\"atze or simplifying assumptions.
The resulting  $B_1(r,-r,0)$ is then compared with the
corresponding ``slice'' obtained from the full kinematic analysis of $B_1(r,p,q)$ mentioned above,
revealing excellent coincidence. This coincidence, in turn, is indicative of an
underlying consonance between elements originating from inherently distinct computational
frameworks, such as the lattice and the SDEs.

The article is organized as follows. In Sec.~\ref{sec:back} we present the  
notation and theoretical ingredients that are relevant for our analysis.
In Sec.~\ref{sec:coupled} we set up and solve the coupled system of SDEs
for the ghost dressing function and the ghost-gluon vertex in general kinematics. 
Next, in Sec.~\ref{sec:cgsoft} we derive and analyze the SDE
for the ghost-gluon vertex in the soft gluon configuration,
comparing our results with those obtained in the previous section.
In addition, we compare the 
strong running coupling obtained from the three-gluon vertex with the one constructed from 
the ghost-gluon vertex, both in the soft gluon configuration.
In Sec.~\ref{sec:conc} we discuss our results and present our conclusions.
Finally, in Appendix~\ref{sec:App_renor} we present useful relations between the Taylor and soft gluon
renormalization schemes, while in Appendix~\ref{sec:App_latt} we discuss the treatment
of finite cut-off effects and lattice scale-setting.


\section{\label{sec:back}Theoretical background}

In this section we 
summarize the main properties of the  two and three-point functions that enter in the nonperturbative determination 
of the ghost-gluon vertex, paying particular attention to the soft gluon limit of
the three-gluon vertex. Note that in the present study  we restrict ourselves to a
\emph{quenched} version of QCD,  \ie a pure Yang-Mills theory with no dynamical quarks.

Throughout this article we work in the Landau gauge,   where the  gluon propagator \mbox{$\Delta^{ab}_{\mu\nu}(q) = -i\delta^{ab} \Delta_{\mu\nu}(q^2)$} assumes the fully transverse form 
\be
    \Delta_{\mu\nu}(q) = \Delta(q^2) P_{\mu\nu}(q) \,,\quad\quad P_{\mu\nu}(q) = g_{\mu\nu} - q_\mu q_\nu/q^2 \,,\quad\quad\, \Delta(q^2)= \Dr(q^2)/q^2.
\label{gluon}    
\ee

As has been firmly established by a variety of large-volume simulations and continuous studies, 
$\Delta(q^2)$ saturates at a finite nonvanishing value, a feature which is widely attributed to the emergence
of a gluonic mass scale~\cite{Cornwall:1981zr,Aguilar:2008xm}.  For later convenience,
the gluon dressing function,  $\Dr(q^2)$, has also been defined in \1eq{gluon}.

 In addition,  we introduce the ghost propagator,    \mbox{$D^{ab}(q^2)=i\delta^{ab}D(q^2)$},  whose 
 dressing function, $F(q^2)$, is given by
\be
D(q^2) =  F(q^2)/q^2 \,,
\label{ghost}
\ee
and is known to saturate at a finite value in the deep infrared~\cite{Boucaud:2008ky,Aguilar:2008xm,Dudal:2008sp,Fischer:2008uz}.

\begin{figure}[h]
 \centering
  \includegraphics[scale=0.65]{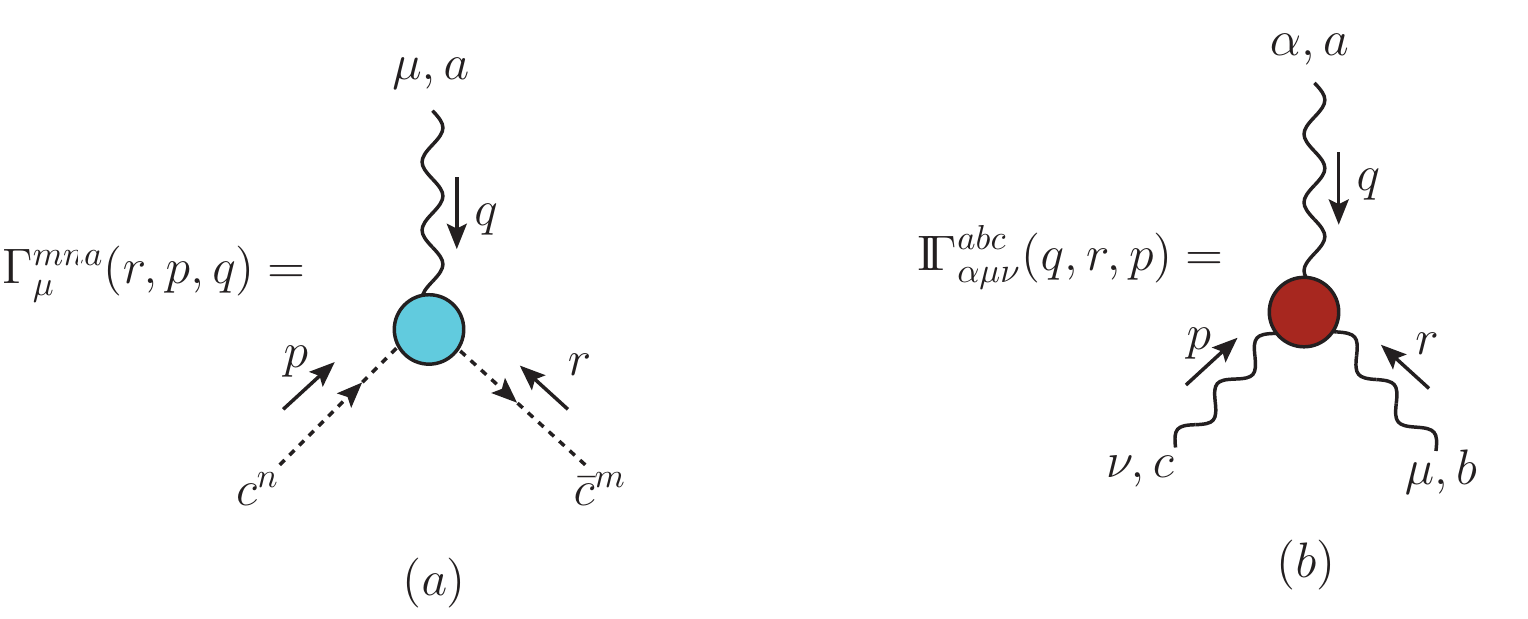}
  \caption{Diagrammatic representation of ($a$)~the ghost-gluon vertex, and ($b$)~the three-gluon vertex, with their respective momenta conventions.  All momenta are incoming,  $q+p+r=0$. }
    \label{vertices}
\end{figure}

Turning to the three-point sector of the theory,
we introduce  the ghost-gluon vertex,
\mbox{$\fatg_\mu^{mna}(r,p,q) = -gf^{mna} \fatg_\mu(r,p,q)$}, 
and the three-gluon vertex, \mbox{$\fatg^{abc}_{\alpha\mu \nu}(q,r,p) =gf^{abc}\fatg_{\alpha\mu \nu}(q,r,p)$},
depicted diagrammatically in the panels ($a$) and ($b$) of Fig.~\ref{vertices}, respectively.

In a series of works~\cite{Aguilar:2011xe,Binosi:2012sj,Ibanez:2012zk,Binosi:2017rwj,Aguilar:2017dco},
the emergence of an infrared finite gluon propagator from the corresponding SDE has been connected 
with certain outstanding nonperturbative features of the fundamental vertices
$\fatg_\mu(r,p,q)$ and $\fatg_{\alpha\mu \nu}(q,r,p)$. 
Specifically, both vertices are composed by two distinct types of terms, according to
\be
\fatg_{\mu}(r,p,q) = \Gamma_{\mu}(r,p,q)+ V_{\mu}(r,p,q),\qquad
\fatg_{\alpha\mu \nu}(q,r,p) = \Gamma_{\alpha\mu \nu}(q,r,p)+ V_{\alpha\mu \nu}(q,r,p)\,.
\label{fatG}
\ee
The terms $V_{\mu}(r,p,q)$ and $V_{\alpha\mu \nu}(q,r,p)$ are purely nonperturbative
and contain longitudinally coupled massless poles; 
when inserted into the SDE of the gluon propagator, 
they trigger the
{\it Schwinger mechanism}~\cite{Schwinger:1962tn,Schwinger:1962tp,Jackiw:1973tr,Eichten:1974et}, 
inducing the infrared finiteness of the gluon propagator.
It is important to emphasize that these terms drop out from transversely projected Green's functions,  or lattice ``observables'',
due to the property\footnote{\ Equivalently, the general tensorial structure of the pole vertices is given by  
\mbox{$V_{\mu}(r,p,q) = \frac{q_\mu}{q^2} A(r,p,q)$} and 
\mbox{$V_{\alpha\mu\nu}(q,r,p) = \frac{q_\alpha}{q^2}  B_{\mu\nu}(q,r,p) +
\frac{r_\mu}{r^2} C_{\alpha\nu}(q,r,p) + \frac{p_\nu}{p^2} D_{\alpha\mu}(q,r,p)$.}} 
\be
\label{eq:transvp}
      {P}_{\mu'}^{\mu}(q) V_{\mu}(r,p,q) =0\,, \qquad\qquad
      {P}_{\alpha'}^{\alpha}(q){P}_{\mu'}^{\mu}(r){P}_{\nu'}^{\nu}(p) V_{\alpha\mu\nu}(q,r,p) = 0 \,.
\ee

On the other hand, the terms $\Gamma_{\mu}(r,p,q)$ and $\Gamma_{\alpha\mu \nu}(q,r,p)$ 
denote the  pole-free components of the two vertices. For large momenta, they
capture the standard perturbative contributions, while in the deep infrared they
may be finite or diverge logarithmically, depending on whether or not 
they are regulated by the nonperturbative gluon mass scale~\cite{Aguilar:2013vaa}.

The most general tensorial decomposition of $\Gamma_\mu(r,p,q)$ can be written as 
\be
 \label{decomp}
    \Gamma_\mu(r,p,q) = B_1(r,p,q)r_\mu + B_2(r,p,q)q_\mu\,,
\ee
where  $B_i(r,p,q)$ are the corresponding form factors.
At  tree-level, \mbox{$\Gamma^{(0)}_\mu =r_\mu $},   and so \mbox{$B_1^{(0)}=1$} and \mbox{$B_2^{(0)}=0$}. In addition,  by virtue of Taylor's theorem~\cite{Taylor:1971ff},  the renormalization constant associated with $\Gamma_\mu(r,p,q)$ is finite.

The vertex $\Gamma_{\alpha\mu \nu}(q,r,p)$ is composed by fourteen linearly independent tensors.
A standard basis, which manifestly reflects the Bose symmetry of $\Gamma_{\alpha\mu \nu}(q,r,p)$,
is the one introduced in~\cite{Ball:1980ax}; see also Eqs.~(3.4) and (3.6) of~\cite{Aguilar:2019jsj}.
Note, however, that the explicit form of the basis will not be required in what follows. 

At tree level, $\Gamma_{\alpha\mu \nu}(q,r,p)$  reduces to the standard expression  
\be
\label{eq:treelevel}
\Gamma_{\!{0}}^{\alpha\mu\nu}(q,r,p) = (q-r)^\nu g^{\alpha\mu} + (r-p)^\alpha g^{\mu\nu} + (p-q)^\mu g^{\nu\alpha}\,.  
\ee

We next turn to the quantity studied in the lattice simulation of~\cite{Aguilar:2021lke},  
\bea
\Ls(r^2) &=&  \frac{\Gamma_0^{\alpha\mu \nu}(q,r,p)
P_{\alpha\alpha'}(q)P_{\mu\mu'}(r)P_{\nu\nu'}(p) \Gamma^{\alpha'\mu'\nu'}(q,r,p)}
{\rule[0cm]{0cm}{0.45cm}\; {\Gamma_0^{\alpha\mu\nu}(q,r,p) P_{\alpha\alpha'}(q)P_{\mu\mu'}(r)P_{\nu\nu'}(p) \Gamma_0^{\alpha'\mu'\nu'}(q,r,p)}}
\rule[0cm]{0cm}{0.5cm} \Bigg|_{\substack{\!\!q\to 0 \\ p\to -r}} \,,  
\label{asymlat}
\eea
where the external legs have been appropriately amputated\footnote{In~\cite{Aguilar:2021lke} and other related lattice works, this quantity has been denominated as the \emph{asymmetric} kinematic limit. Here
  we find it more appropriate to employ the term ``soft gluon limit''.}. Note that the starting expression involves the full vertex 
$\fatg^{\alpha'\mu'\nu'}(q,r,p)$, which, by virtue of \1eq{eq:transvp}, is reduced to  $\Gamma^{\alpha'\mu'\nu'}(q,r,p)$,  
\ie the term $V^{\alpha'\mu'\nu'}(q,r,p)$ associated with the poles drops out in its entirety. 

Now, in the limit of interest, namely $q \to 0$, the tensorial structure of 
the three-gluon vertex is considerably simplified, given by
\be
\Gamma^{\alpha\mu\nu}(0,r,-r) = 2 {\cal A}_1(r^2) \,r^\alpha g^{\mu\nu} + {\cal A}_2(r^2)\, (r^\mu g^{\alpha\nu} + \,r^\nu g^{\alpha\mu})
+ {\cal A}_3(r^2)\, r^\alpha r^\mu r^\nu  \,.
\label{Gsoft}
\ee
At tree level,
\be
\Gamma_0^{\alpha\mu\nu}(0,r,-r) = 2 \,r^\alpha g^{\mu\nu} - (r^\mu g^{\alpha\nu} + \,r^\nu g^{\alpha\mu})\,,
\label{Gsoft0}
\ee
which, in the notation of \1eq{Gsoft}, means that ${\cal A}_1^{(0)}(r^2)= 1$,  ${\cal A}_2^{(0)}(r^2)=-1$,
and ${\cal A}_3^{(0)}(r^2)= 0$.

Then, the numerator and denominator of the fraction on the r.h.s. of \1eq{asymlat}, to be denoted
by ${\cal N}$ and ${\cal D}$, respectively, become 
\be
{\cal N} = 4 (d-1) [r^2 - (q\cdot r)^2/q^2] {\cal A}_1(r^2)\,, \qquad  {\cal D} = 4 (d-1) [r^2 - (q\cdot r)^2/q^2] \,.
\label{NandD}
\ee
Thus, the path-dependent contribution contained in the square bracket drops out when forming the ratio ${\cal N}/{\cal D}$,
and \1eq{asymlat} yields simply
\be
\Ls(r^2) =  {\cal A}_1(r^2) \,.
\label{LisB}
\ee
Combining \2eqs{Gsoft0}{LisB},  it is immediate to derive one of the key relations of this work, namely 
\be
\label{GammaLasym}
P_{\mu\mu'}(r) P_{\nu\nu'}(r) \Gamma^{\alpha\mu\nu}(0,r,-r) = 2 \Ls(r^2) r^{\alpha}P_{\mu'\nu'}(r) \,.
\ee

\section{\label{sec:coupled} The system of coupled SDEs}

In this section, we set up and solve  
the system of coupled SDEs that governs the ghost dressing function and the ghost-gluon vertex for general space-like momenta. 
The  external ingredients employed are a fit of the lattice data for the gluon propagator, and certain form factors
of the three-gluon vertex (in general kinematics), obtained from the nonperturbative Ball-Chiu construction of~\cite{Aguilar:2019jsj}.

\begin{figure}[b]
    \centering
        \vspace{-0.5cm}
    \includegraphics[scale=0.7]{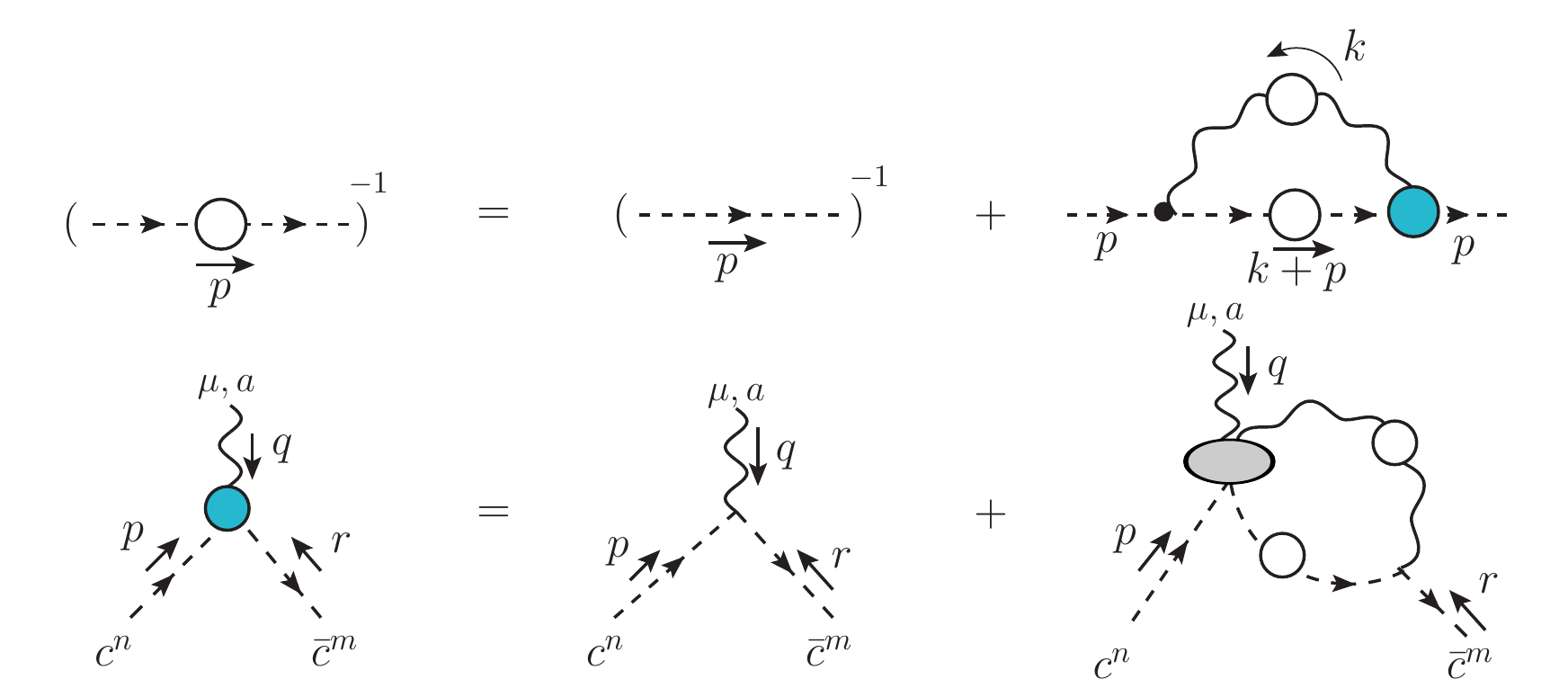}
    \vspace{-0.5cm}
    \caption{The SDEs for the ghost propagator and the ghost-gluon  vertex (upper and lower panels, respectively). 
      The white circles represent the full gluon and ghost propagators,  while the blue ones
 denote the full ghost-gluon vertex.  The gray ellipse indicates the ``one-particle reducible'' four-point ghost-gluon kernel.}    
    \label{fig:coupsys}
\end{figure}

\subsection{The ghost gap equation and ghost-gluon SDE}

Our starting point is the SDE for the ghost propagator, whose diagrammatic representation is shown in the upper panel of Fig.~\ref{fig:coupsys}.
When expressed in terms of the ghost dressing function, this SDE acquires the standard form known in the literature, namely 
\be
F ^{-1}(p^2) =  Z_{\rm c} + \Sigma(p^2)\,,
\label{sde_ghost2}
\ee
with
\be
\Sigma(p^2) =  ig^2C_{\rm A}Z_1\!\! \int_k f(k,p)B_1(-p,k+p,-k)\Delta(k)D(k+p)\,; \qquad f(k,p):= 1 -\frac{(k\cdot p)^2}{k^2p^2}\,.
\label{sde_ghost}
\ee
In the above equation, $C_\mathrm{A}$ is the Casimir eigenvalue of the adjoint representation [$N$ for $SU(N)$],
while $Z_{\rm c}$ and $Z_1$ are the  renormalization constants of $D(p^2)$ and  $\Gamma^\mu (r,p,q)$, respectively  [see Eq.~\eqref{renorm}].
 In addition,   we have introduced the integral measure 
\begin{equation}
    \int_k := \frac{1}{(2\pi)^4} \int\!\! \dd[4]{k} \,,
\end{equation}
where the presence of a symmetry-preserving regularization scheme is implicitly understood.

In this analysis, the renormalization is implemented within the well-known variant of the momentum subtraction
(MOM) scheme known as ``Taylor scheme''~\cite{Boucaud:2008gn,Boucaud:2011eh}\footnote{In the literature this scheme is also known as  minimal momentum subtraction (MiniMOM) scheme~\cite{vonSmekal:2009ae}, 
and has been employed for a recent determination of $\alpha_{\ols{\srm{MS}}}$ from unquenched lattice simulations~\cite{Zafeiropoulos:2019flq}, consistent with the experimental \emph{world average.}}, 
which fixes the (finite) vertex renormalization constant at the special value $Z_1=1$.
As for $Z_{\rm c}$, its value is fixed by the standard MOM requirement 
$F ^{-1}(\mu^2) = 1$, where $\mu$ is the renormalization scale.

Implementing this condition at the level of 
\1eq{sde_ghost2} yields
\be
Z_{\rm c} = 1 - \Sigma (\mu^2)\,,
\ee
and \1eq{sde_ghost2} may be cast in the form
\be
F ^{-1}(p^2) =  1 + \Sigma(p^2) - \Sigma (\mu^2)\,.
\label{sde_ghostks}
\ee

\begin{figure}[t]
    \centering
    \includegraphics[scale=0.7]{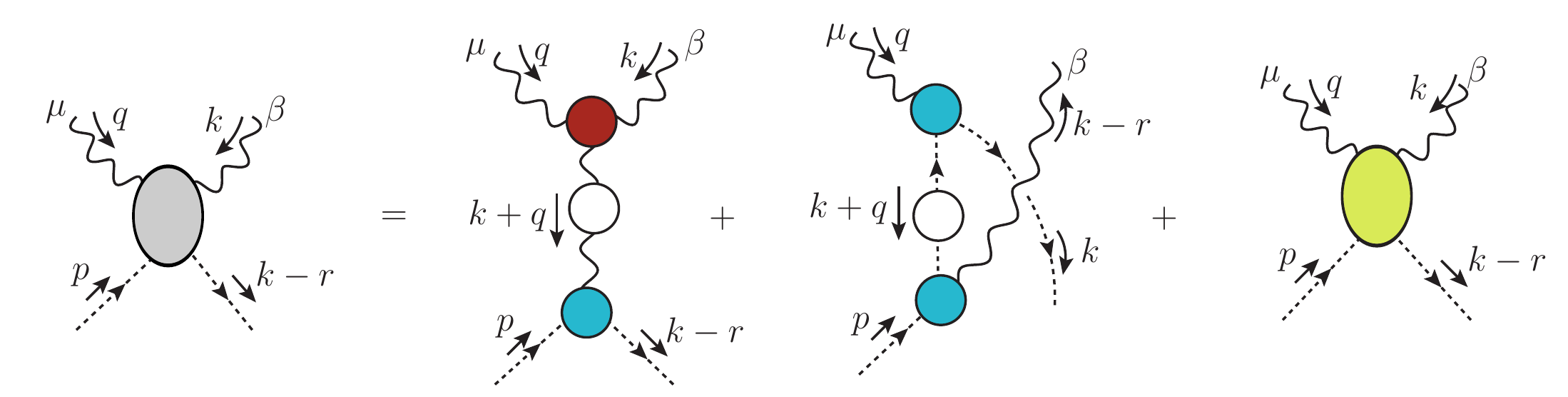}
    \caption{The skeleton  expansion of the ``one particle reducible'' four-point ghost-gluon kernel.
     Only the first two terms will be considered in our analysis.}
    \label{fig:diagvert}
\end{figure}

We next turn to the SDE for the ghost-gluon vertex, shown diagrammatically in the lower panel of Fig.~\ref{fig:coupsys}.
In the present work,  we will consider the so-called ``one-loop dressed'' approximation of this SDE, which
corresponds to keeping only the first two terms in the skeleton expansion of the SDE kernel, shown in Fig.~\ref{fig:diagvert}.
Note that the omitted set of contributions is captured by the one-particle irreducible four-point function, represented by the yellow ellipse,
whose dynamics has been studied in detail in~\cite{Huber:2017txg,Huber:2018ned}. As was shown there, this subset of corrections
is clearly subleading, affecting the ghost-gluon vertex by a mere $2\%$. 
It is therefore expected that the above truncation should provide a quantitatively accurate description of
the infrared behavior of the ghost-gluon vertex [see also the corresponding discussion in Sec.~\ref{sec:conc}]. 

Thus,  the expression for the SDE for the ghost-gluon vertex in the Taylor scheme can  be schematically written as 
\be
 \label{rvgg}
 \Gamma_\mu(r,p,q) = r_\mu - \frac{i}{2}g^2 C_\mathrm{A} [a_{\mu}(r,p,q) - b_{\mu}(r,p,q)] ,
\ee
with
\begin{align}
    a_\mu(r,p,q) &= r_\rho \int_k \Delta^{\rho\sigma}(k) \fatg_{\mu\sigma\alpha}(q,k,-t) \Delta^{\alpha\beta}(t) \Gamma_\beta(-\ell,p,t) D(\ell) \,,  \nonumber\\
    b_\mu (r,p,q)&= r_\alpha \int_k \Delta^{\alpha\beta}(\ell) \Gamma_\beta(t,p,-\ell) D(t) \fatg_\mu(k,-t,q) D(k)  \,, 
\label{d1d2}
\end{align}
where $\ell := k-r$  and $t :=k+q$. Note that we have employed the first of the two relations in \1eq{eq:transvp} in order to eliminate
the terms $V_{\mu}(r,p,q)$ from the ghost-gluon vertices that are contracted by a transverse gluon propagator
(Landau gauge). 


In order to  isolate  the contribution of the form factor $B_1(r,p,q)$, defined in \1eq{decomp}, we contract \1eq{rvgg}
by the projector~\cite{Aguilar:2013xqa}
\be 
 \varepsilon^\mu(r,q) = \frac{q^2 r^\mu - q^\mu ( q\cdot r )}{ h(r,q) }\,, \qquad h(q,r) = q^2 r^2 - (q\cdot r)^2 \,.
\label{projector}
\ee
An immediate consequence of this contraction and the property \1eq{eq:transvp} is that 
\bea 
\fatg_{\mu\sigma\alpha}(q,k,-t)P^{\rho\sigma}(k) P^{\alpha\beta}(t)
&\to & \Gamma_{\mu\sigma\alpha}(q,k,-t)P^{\rho\sigma}(k) P^{\alpha\beta}(t)\,,
\nonumber\\
\fatg_\mu(k,-t,q) &\to & \Gamma_\mu(k,-t,q)\,,
\label{subst}
\eea
\ie the terms associated with the nonperturbative poles are annihilated, and we are only left with the
pole-free components of the two vertices.


The next step is to carry out in the expressions of \1eq{d1d2} the substitution
\begin{align}
B_1(-\ell, p, t) \to  & \frac{1}{2}\left[ B_1(-\ell, p, t) + B_1(r, \ell, - k) \right]\,,  \nonumber \\
B_1(t, p, -\ell) \to & \frac{1}{2}\left[ B_1(t, p, -\ell) + B_1(r, -k, \ell) \right]\,,
\label{sym_sub}
\end{align}
in order to restore the symmetry of $B_1(r,p,q)$ with respect to the interchange of the ghost and antighost momenta,
which has been compromised by the truncation of the SDE~\cite{Aguilar:2018csq}.

In addition, the structure of the three-gluon vertex entering in $a_\mu(r,p,q)$ is approximated
by retaining only the tensorial structures with a nonvanishing tree-level limit. Specifically,
in the notation of~\cite{Aguilar:2019jsj}, we set 
\be
\label{eq:vtensortree}
\Gamma^{\alpha\mu\nu}(q,r,p) \approx (q-r)^\nu g^{\alpha\mu}X_1(q,r,p) + (r-p)^\alpha g^{\mu\nu} X_4(q,r,p)
+ (p-q)^\mu g^{\nu\alpha} X_7(q,r,p)\,,  
\ee
where, due to the Bose symmetry of $\Gamma^{\alpha\mu\nu}(q,r,p)$, we have \mbox{$X_1(q,r,p) = X_4(p,q,r) = X_7(r,p,q)$}.

Thus, we  arrive at (Minkowski space)  
\be
B_1(r,p,q) = 1 -  \frac{i}{2}g^2 C_\mathrm{A} \left[a(r,p,q)  - b(r,p,q) \right]\,,
 \label{B1general}
\ee
with
\be
a(r,p,q) = \int_k  {\mathcal K}_1(k,r,q){\mathcal N}_1 (k,r,q)\,,  \qquad \quad  b(r,p,q) =\int_k  {\mathcal K}_2(k,r,q)  {\mathcal N}_2 (k,r,q) \,,  
\label{d12kernels}
 \ee
where
\bea
{\mathcal K}_1(k,r,q) &=&   \frac{\Delta(k^2) \Delta(t^2) F(\ell^2)} { k^2 \ell^2 t^2 h(q,r) } \left[ B_1(-\ell, p, t) + B_1(r, \ell, - k) \right]\,,  \nonumber  \\
{\mathcal K}_2 (k,r,q)&=&  \frac{F(k^2) \Delta( \ell^2 ) F(t^2) }{ 2\, k^2 \ell^2 t^2 h(q,r) }\left[ B_1(t, p, -\ell) + B_1(r, -k, \ell) \right]B_1(k,-t,q) \,, 
\label{kb1}
\eea
and 
\bea
{\mathcal N}_1 &=&  a_1X_1(k, t, q) +a_4X_4(k, t, q)+a_7X_7(k, t, q)\,,  \\ 
{\mathcal N}_2&=&   [ q^2 ( k \cdot r ) - ( k\cdot q )(q\cdot r ) ] [ (k\cdot r)(q\cdot r) + (k\cdot q )(k\cdot r) - r^2 (k\cdot q ) - k^2(q\cdot r ) - h(k,r) ]\,.   \nonumber 
\eea
The coefficients $a_i$ are given by
\begin{align}
a_1 =&  \left[ q^2 ( k\cdot r ) - ( k\cdot q ) (q\cdot r) \right] \left\lbrace k^2 \left[ ( k\cdot q ) (k\cdot r) - ( k\cdot q )(q\cdot r)-2 r^2( k\cdot q ) \right. \right. \nonumber\\
& \left.\left. + (k\cdot r)( q\cdot r) + (k\cdot r)^2 - h(q,r) \right] - k^4 \left[ (q\cdot r ) + r^2 \right] + ( k\cdot r ) \left[ q^2 (k\cdot r) + (k\cdot q) (k\cdot r) \right.\right. \nonumber\\
& \left.\left. - (k\cdot q)(q\cdot r) + (k\cdot q)^2\right] \right\rbrace \,, \nonumber \\
a_4 =&  \left[ k^2 ( q\cdot r ) - ( k\cdot q ) ( k\cdot r ) \right] \left\lbrace \left(k^2+q^2\right) h(q,r) - q^2 ( k\cdot r ) \left[ q^2 + ( q\cdot r ) \right] + (k\cdot q)^2 (q\cdot r) \right.\nonumber\\
& \left. + ( k\cdot q )\left[ ( k\cdot r ) (q\cdot r) - q^2 ( k\cdot r ) + q^2 (q\cdot r) + 2 q^2 r^2 - (q\cdot r)^2 \right] - q^2 (k\cdot r)^2 \right\rbrace \,, \nonumber\\
a_7 =& \left\lbrace q^2 ( k\cdot r ) - k^2 \left[ q^2 + ( q\cdot r ) \right] + ( k\cdot q ) [ ( k\cdot r ) - ( q\cdot r ) ] + ( k\cdot q)^2 \right\rbrace \nonumber\\
& \times \left[ k^2 h(q,r) - q^2 (k\cdot r)^2 + ( k\cdot q )( k\cdot r )( q\cdot r ) \right]\,. 
\label{ai_Mink}
\end{align} 

\subsection{\label{sec:ngene} Numerical analysis}

In order to proceed with the numerical solution, the system of integral equations formed by
\2eqs{sde_ghostks}{B1general} must be passed to Euclidean space, 
following standard conventions (see,  \eg Eq.~(5.1) of~\cite{Aguilar:2018csq}) and employing spherical coordinates for the
final treatment.

Then, appropriate inputs for the gluon propagator, $\Delta(q^2)$,  
and the form factors $X_{1,4,7}(q,r,p)$ of the three-gluon vertex must be furnished.

\begin{figure}[t]
\begin{minipage}[b]{0.45\linewidth}
\centering
\hspace{-1.0cm}
\includegraphics[scale=0.24]{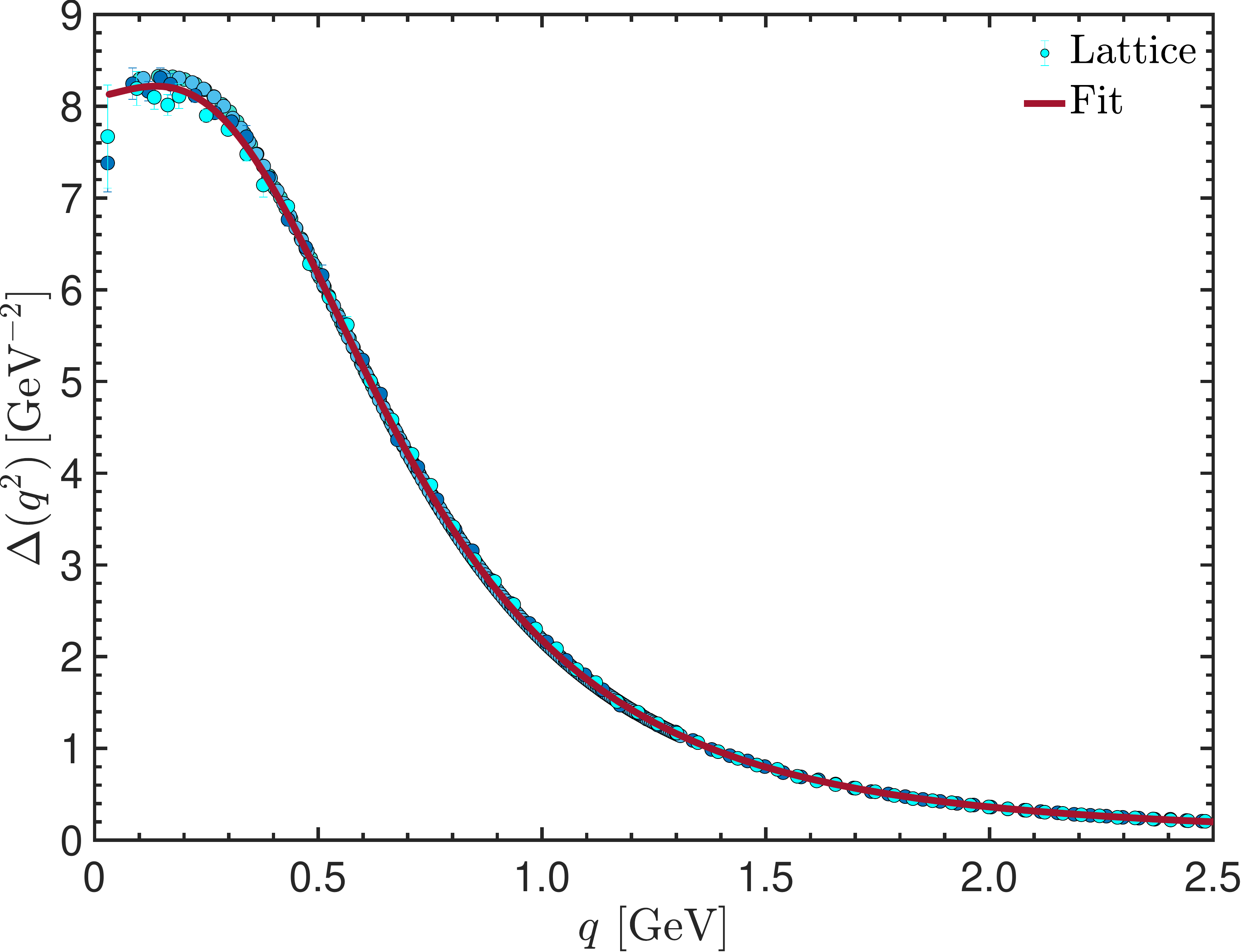}
\end{minipage}
\hspace{0.25cm}
\begin{minipage}[b]{0.45\linewidth}
\includegraphics[scale=0.9]{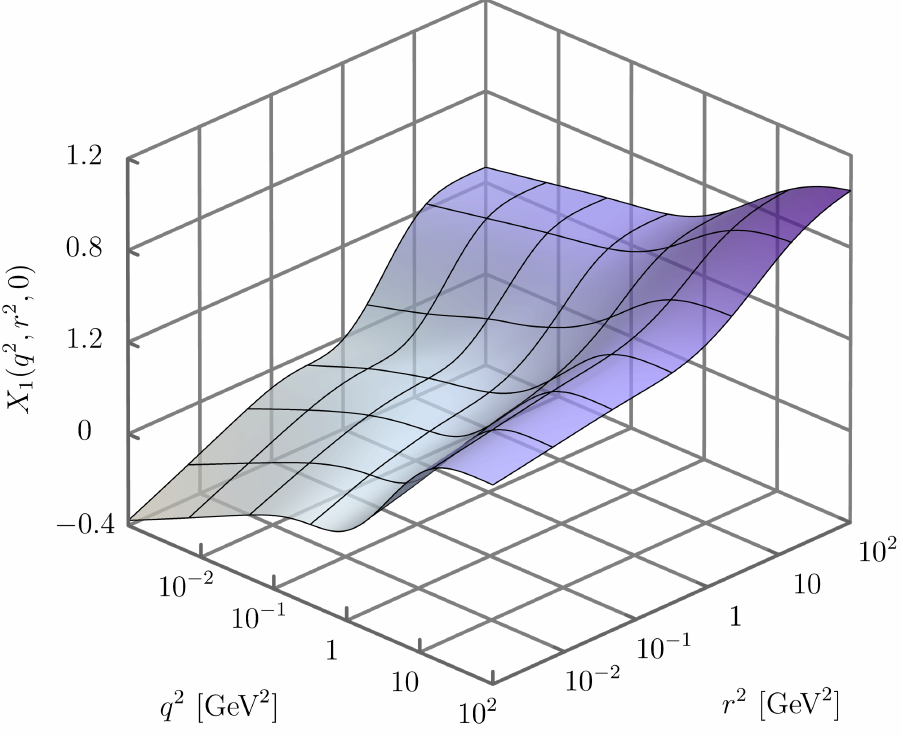}
\end{minipage}
\caption{Left panel: Lattice data for the gluon propagator,  $\Delta(q^2)$,  after performing the continuum extrapolation of~\cite{Boucaud:2018xup} to the data set of~\cite{Bogolubsky:2009dc},   together with the corresponding fit  given by Eq.~\eqref{gluonfit}.   The gluon propagator is renormalized  at \mbox{$\mu = 4.3$ GeV}.  Right panel: A representative case of the three-gluon form factor  $X_1(q^2,r^2, \phi)$  for a fixed value of the angle, $\phi=0$.} 
\label{fig:X1}
\end{figure}

For the gluon propagator we employ a fit for the results obtained after a  reanalysis of the lattice data of~\cite{Bogolubsky:2009dc},
following the procedure put forth in~\cite{Boucaud:2017ksi,Boucaud:2018xup}, in order to cure volume and discretization artifacts, 
see Appendix~\ref{sec:App_latt} for details. Specifically, the resulting $\Delta(q^2)$ is shown in the
left panel of Fig.~\ref{fig:X1}, together with the numerical fit given by Eq.~\eqref{gluonfit}.

\begin{figure}[t]
\begin{minipage}[b]{0.45\linewidth}
\centering
\hspace{-1.0cm}
\includegraphics[scale=0.26]{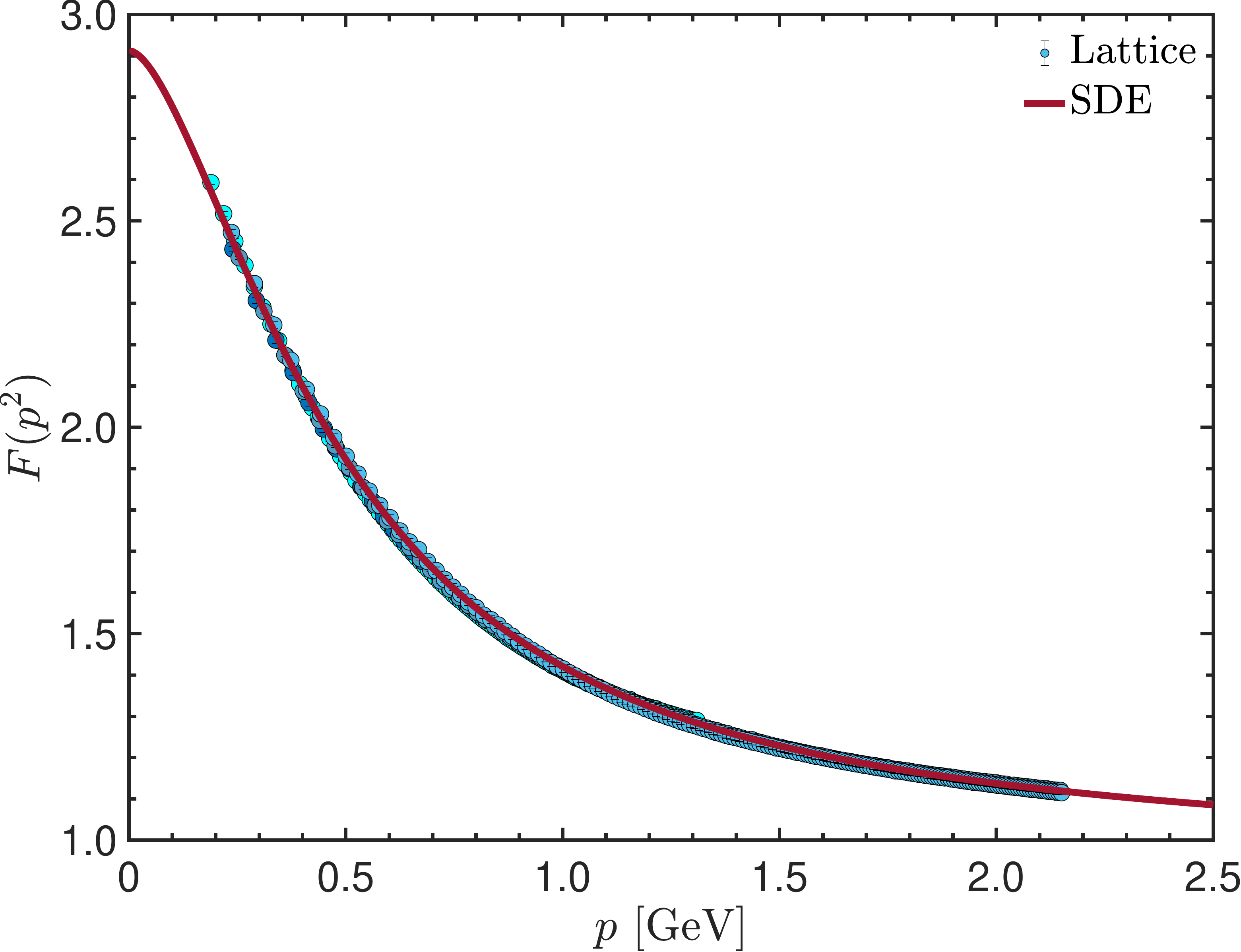}
\end{minipage}
\hspace{0.25cm}
\begin{minipage}[b]{0.45\linewidth}
\vspace{0.5cm}
\includegraphics[scale=0.9]{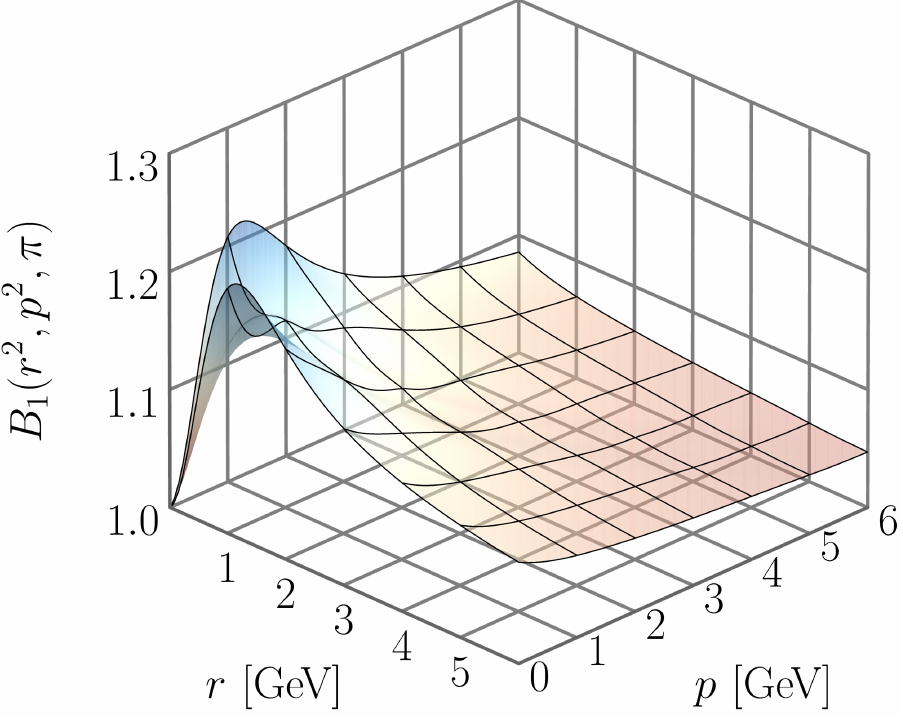}
\end{minipage}
\caption{ Left panel:  The numerical solution for the  ghost dressing function,  $F(p^2)$,  (red continuous line) compared with the lattice data of~\cite{Boucaud:2018xup}.   Right panel:  The form factor  $B_1(r^2,p^2,\theta_1 )$  for a fixed value of the angle $\theta_1 =\pi$,  obtained as solution of the coupled system of  \2eqs{sde_ghost2}{B1general} when  \mbox{$\alpha_s(\mu)= 0.244$}.  }
\label{fig:rescoupled}
\end{figure}

For the determination of the form factors $X_{1,4,7}(q,r,p)$,  we follow the nonperturbative version of the Ball-Chiu construction
developed in~\cite{Aguilar:2019jsj}. The general idea of the method is based on reconstructing the longitudinal form factors
of the three-gluon vertex, such as $X_{1,4,7}(q,r,p)$, from the set of STIs that $\Gamma_{\alpha\mu \nu}(q,r,p)$ satisfies.
This procedures allows us to express $X_{1,4,7}(q,r,p)$ in terms of the ghost dressing function, the ``kinetic'' part of the
gluon propagator, and three of the form factors of the ghost-gluon kernel.  A representative case of $X_1(q^2,r^2,\phi=0)$  
is shown in the right panel of Fig.~\ref{fig:X1}, where $\phi$ is the angle formed between the momenta $q$ and $r$. 
Note that the form factor deviates markedly from unity, displaying clearly what is known in the literature
as ``infrared suppression''~\cite{Aguilar:2013vaa,Athenodorou:2016oyh,Boucaud:2017obn,Blum:2015lsa,Corell:2018yil,Aguilar:2019jsj}.

With the inputs introduced above, the coupled system is solved numerically by an iterative process.  The external momenta $r^2$ and $p^2$ are distributed on a logarithmic grid, with $96$ points in the interval \mbox{$[5\times10^{-5},10^4]$ GeV$^2$},  whereas the angle between them,  $\theta_1$,   is uniformly distributed in \mbox{$[0,\pi]$} with 19 points. The interpolations in three variables, needed for evaluating the $X_i$ and the $B_1$, are performed with B-splines~\cite{de2001practical}, and the triple integrals are computed with a Gauss-Kronrod method~\cite{Berntsen:1991:ADA:210232.210234}.

In Fig.~\ref{fig:rescoupled},  we show the numerical results for  $F(p^2)$ and $B_1(r^2,p^2, \theta_1)$,  
obtained from the solution of the coupled system.
We emphasize that the renormalization point has been fixed at \mbox{$\mu= 4.3$ GeV},
which coincides with the highest value of the momentum
accessible by the lattice simulation of~\cite{Bogolubsky:2009dc}.
In particular,   
one can observe that when the gauge coupling assumes the value \mbox{$\alpha_s(\mu):=g^2(\mu)/4\pi = 0.244$},
the solution of the system yields a $F(p^2)$ that is in outstanding agreement with the  ghost dressing data of~\cite{Boucaud:2018xup} (left panel),  which were properly extrapolated to the physical continuum limit, 
as  explained in Appendix~\ref{sec:App_latt}. 

\begin{figure}[t]
\begin{minipage}[b]{0.45\linewidth}
\centering
\hspace{-1.0cm}
\includegraphics[scale=0.9]{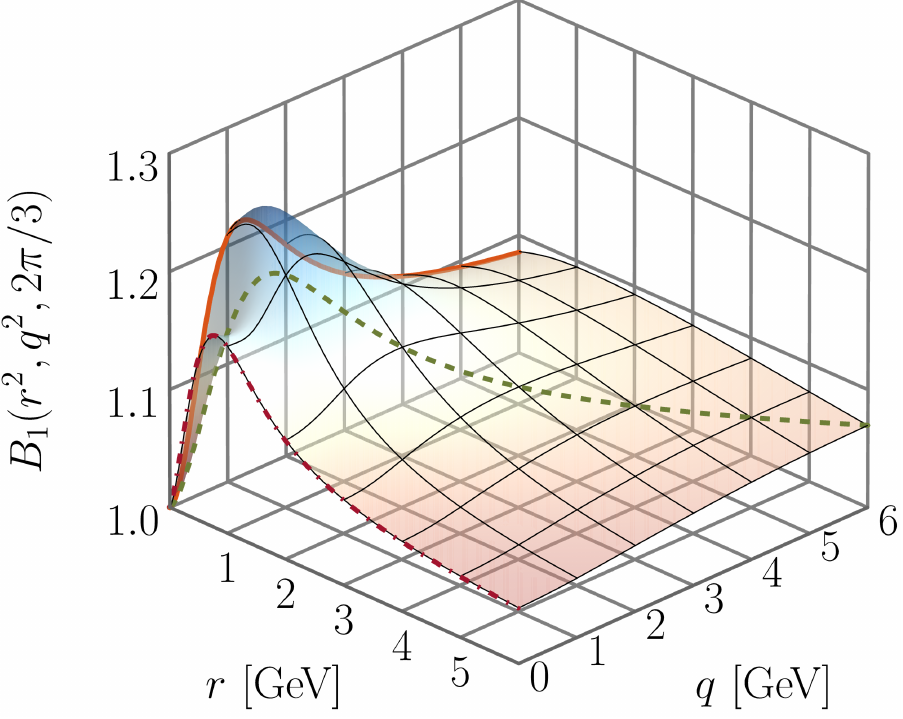}
\end{minipage}
\hspace{0.25cm}
\begin{minipage}[b]{0.45\linewidth}
\includegraphics[scale=0.24]{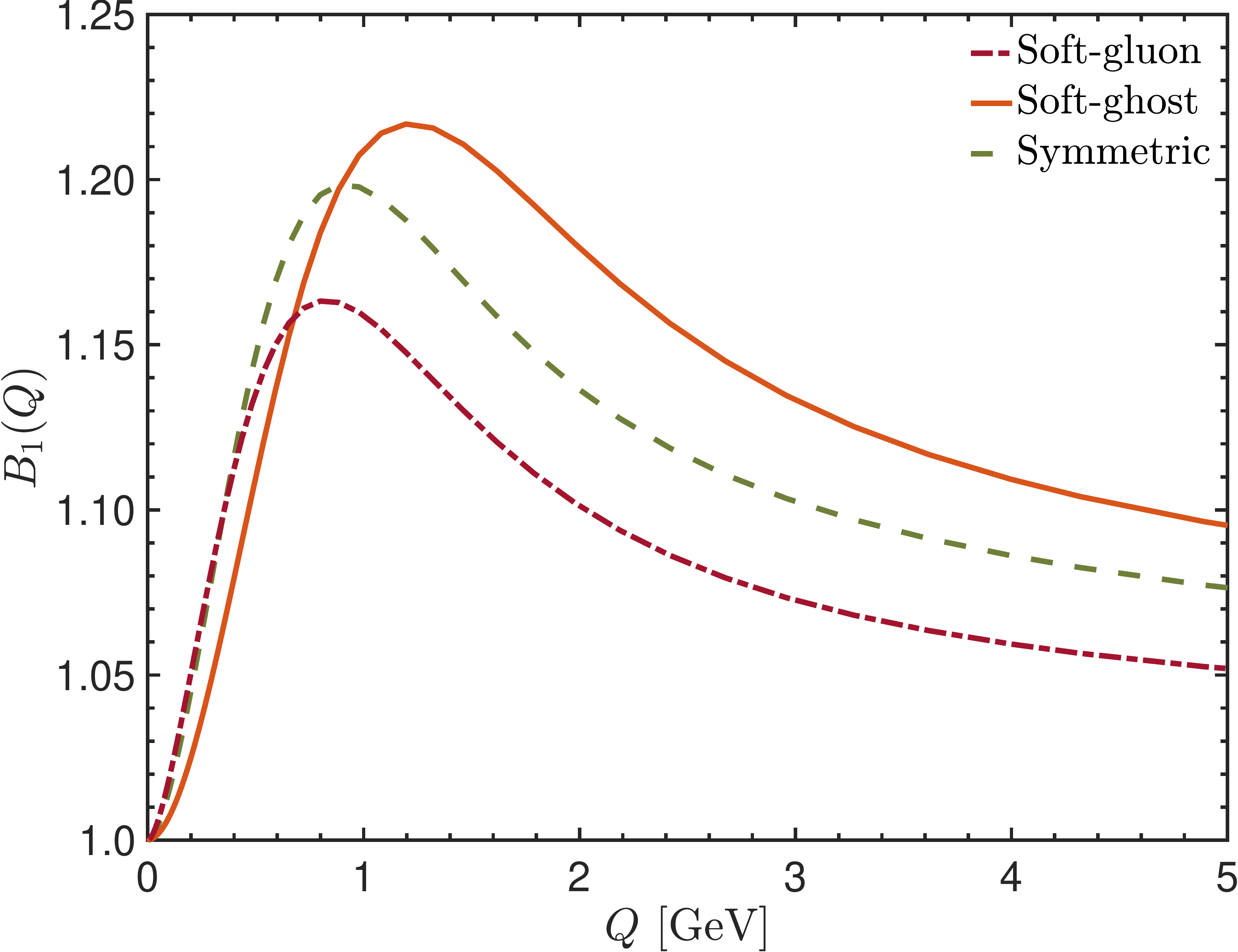}
\end{minipage}
\caption{Left panel: The form factor $B_1(r^2,q^2,\theta_2)$ plotted as function of the momenta of antighost,  $r$,   and  gluon,  $q$,   for a fixed value of the angle,  $\theta_2=\pi$.
On the  3-D surface, three curves are highlighted, representing
the soft gluon (red dot-dashed), soft ghost (orange continuous), and symmetric (green dashed) kinematic limits.  Right panel: Direct comparison of the three special configurations (2-D projections)
identified in the left panel. } 
\label{fig:conf}
\end{figure}

Moreover,  in the right panel of   Fig.~\ref{fig:conf},    one can see that the solution for  $B_1(r^2,p^2, \theta_1)$  is symmetric with respect to  the diagonal plane defined by the condition $r=p$.
This is a direct consequence of  the ghost-antighost symmetry, and becomes manifest only when $B_1$ is plotted  as a
function of the momenta $r$ and $p$.

We next explore certain special kinematic limits of $B_1$. To that end,
we choose  $r$ and $q$ as our reference momenta (antighost and gluon, respectively), denoting by 
$\theta_2$ the angle between them.  In the left panel of Fig.~\ref{fig:conf} we plot the corresponding
3-D plot, for the special value $\theta_2=2\pi/3$; this choice for the angle  
is particularly convenient,  because  one can identify  
on a unique  3-D surface the following three kinematic limits:

{\it(i)} The  \emph{soft gluon limit},  obtained by setting 
$q = 0$; then,  the momenta $r$ and $p$ have the same magnitude, \mbox{$|p|=|r|=|Q|$}, and are anti-parallel, \ie  $\theta_1=\pi$.
This kinematic configuration is represented by the red dot-dashed curve on the 3-D plot of Fig.~\ref{fig:conf}.

 {\it(ii)} 
 The \emph{soft (anti)ghost limit}, in which $r= 0$  and 
the momenta \mbox{$|q|=|p|=|Q|$}; evidently, $|r||q|\cos\theta_2= 0$,
and any dependence on the angle $\theta_2$ is washed out. This kinematic limit is represented by the orange continuous curve 
on the 3-D plot of Fig.~\ref{fig:conf}.

{\it(iii)} The \emph{totally symmetric limit},  defined by \mbox{$q^2 = p^2= r^2 = Q^2$}; with the
scalar products given by 
  \mbox{$(q\cdot p) = (q\cdot r) = (p\cdot r) = -\frac{1}{2}Q^2$},   and the angles  \mbox{$ \widehat{rp} =  \widehat{rq} =  \widehat{qp} =2\pi/3$},
represented by the green dashed curve  on the 3-D plot of Fig.~\ref{fig:conf}.

The three 2-D projections described above are plotted together in the right panel of Fig.~\ref{fig:conf},
with all their corresponding momenta denoted by $Q$. As we can see, all cases display a peak around the same region of momenta, \ie \mbox{$(0.8-1.2)$ GeV},  with moderate differences in their heights.  In addition,  in the deep infrared, all curves recover
the result $B_1(0,0,0) = 1$.

\section{\label{sec:cgsoft}Ghost-gluon vertex in the soft gluon configuration}

In this Section we implement the soft gluon limit,  \ie ($q \rightarrow 0$),  directly at the level of the
SDE for the ghost-gluon vertex, which permits us to 
use the lattice data for $\Ls(q^2)$~\cite{Aguilar:2021lke}\footnote{In~\cite{Aguilar:2021lke}, the lattice result for $\Ls(q^2)$
  has been reproduced particularly well by means of the STI-based construction of~\cite{Aguilar:2019jsj}. Nonetheless, in the present analysis we
employ directly the best fit to the lattice data, for achieving the highest possible accuracy.}
in the treatment of the resulting
integral equation.

The basic observation is that, in the soft gluon limit,  
the term  \mbox{$P^{\rho\sigma}(k)  \Gamma_{\mu\sigma\alpha}(q,k,-t) P^{\alpha\beta}(t)$} appearing 
inside the $a_\mu(r,p,q)$ of \1eq{d1d2} becomes simply  
\be
P^{\rho\sigma}(k) P^{\alpha\beta}(t) \Gamma_{\mu\sigma\alpha}(q,k,-t) \xrightarrow[\text{}]{\text{$q\to 0$}}
P^{\rho\sigma}(k)P^{\alpha\beta}(k) \Gamma_{\sigma\alpha\mu}(0,k,-k) = 2 \Ls(k^2) k_{\mu}P^{\rho\beta}(k)\,,  
\ee
where in the last step \1eq{GammaLasym} was used. 

Note, however, that a final subtlety prevents the immediate use of the lattice results 
for $\Ls(k^2)$ into \1eq{d1d2}. Specifically, the renormalization employed in the lattice  
analysis of~\cite{Aguilar:2021lke} is the ``soft gluon scheme'', which differs from the Taylor scheme
used in the derivation of the system of coupled SDEs. As a result, the lattice data must
undergo a finite renormalization, $\tilde{z}_{3}$, which will convert
them from one scheme to the other, according to \1eq{asytaylor}.  

Then, it is straightforward to implement the soft gluon limit at the level of \1eq{d1d2}.
Using the short-hand notation \mbox{$B_1(k^2):=B_1(k,-k,0)$}, we arrive at 
\begin{align}
  B_1(r^2) & = 1 - \frac{ig^2C_\mathrm{A}}{\tilde{z}_3} \int_k F(\ell^2) \Delta^2(k^2)
  f(k,r)\frac{(k\vdot r)}{\ell^2} B_1(-\ell,-r,k)\Ls(k^2) \nonumber \\
    & + \frac{i}{2} g^2C_\mathrm{A} \int_k F^2(k^2) \Delta(\ell^2)  f(k,r)\frac{(k\vdot r)}{k^2\ell^2} B_1(k,-r,-\ell) B_1(k^2) \,, 
\label{B1exp}    
\end{align} 
 where the function $f(k,r)$ has been defined in \1eq{sde_ghost}.

As a final step, \1eq{B1exp} will be converted to Euclidean space (spherical coordinates),
using standard transformation rules.  Defining 
 \begin{equation}
    k^2:= y; \qquad r^2:=x; \qquad \ell^2:= z; \qquad k\vdot r\equiv \sqrt{xy}\cos{\theta}; \qquad \ell \vdot r\equiv \sqrt{xz}\cos{\varphi};
 \end{equation}
and setting 
\bea 
B_1 (\ell ,-r, k) \to  B_1(z,x,\varphi)\,,    \qquad    B_1 (k ,-r, -\ell) \to  B_1(y,x,\pi -\theta)\,,   
\label{ffeu}
\eea
we arrive at 
\bea
B_1(x) & =& 1 +\frac {C_\mathrm{A}\alpha_s}{2\pi^2 \,\tilde{z}_{3}} \int_0^\infty\!\!\! \dd{y} y\sqrt{xy}\,  \Ls(y) \Delta^2(y) \int_0^\pi\!\!\! \dd{\theta} \sin^4 \theta \cos{\theta} B_1(z,x,\varphi) \, z^{-1} F(z)\nonumber\\   
      & +& \frac{C_\mathrm{A}\alpha_s}{4\pi^2}\int^\infty_0\!\!\! \dd{y} \,\sqrt{xy} F^2(y) B_1(y) \int_0^\pi \!\!\! \dd{\theta}
 \sin^4\theta \cos{\theta} B_1(y,x,\pi-\theta)\, z^{-1} \Delta(z) \,, 
\label{FinalEq}
\eea
where we have that $\cos{\varphi}= \sqrt{y/z} \cos\theta - \sqrt{x/z}$.

\1eq{FinalEq} will be solved numerically, through an iterative procedure, using the
following external inputs.

\begin{figure}[t]
\centering
\includegraphics[scale=0.28]{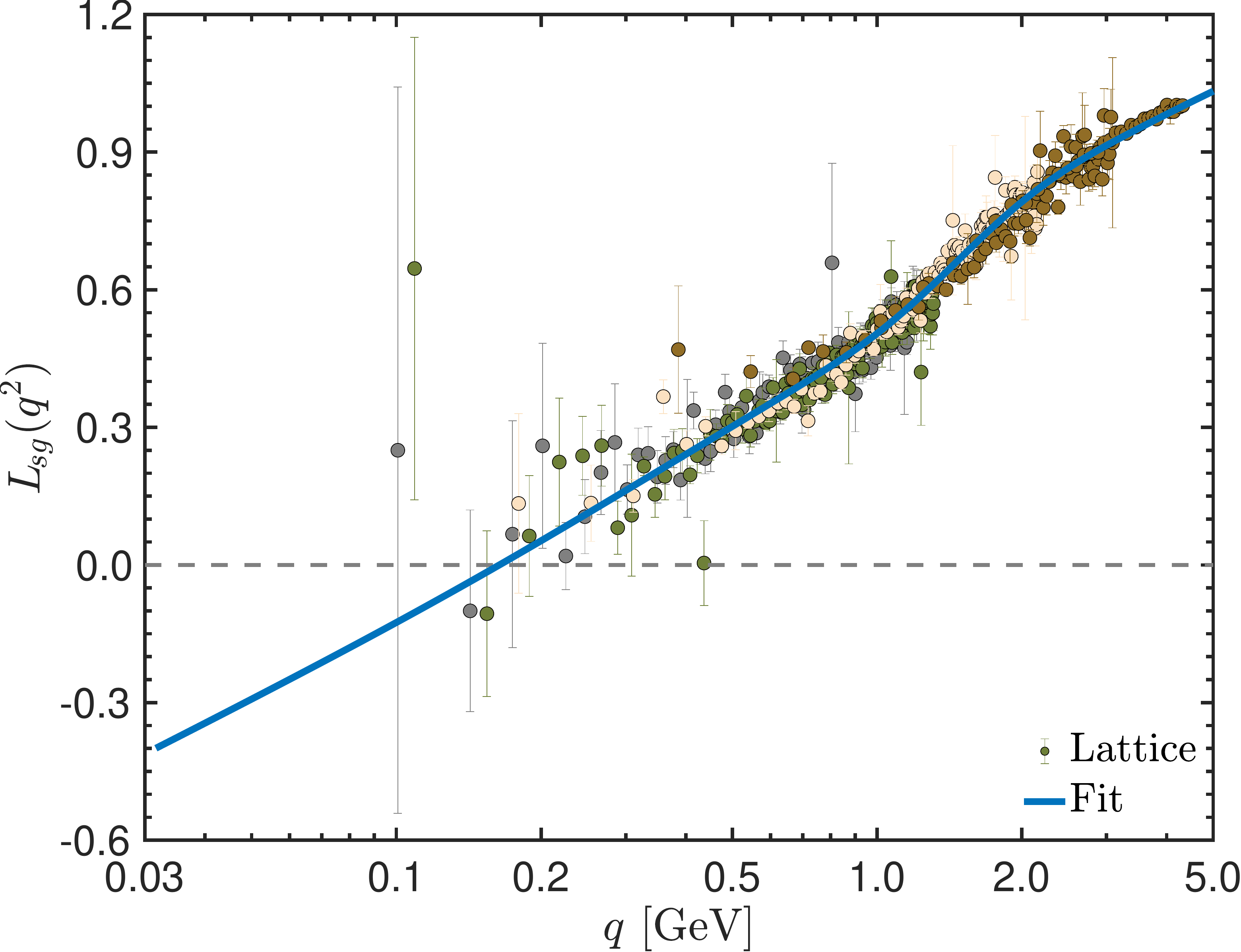}
\caption{The  lattice data for $\Ls(q^2)$ (circles)  from~\cite{Aguilar:2021lke},  together with the fit given by 
Eq.~\eqref{Lasymfit} (blue continuous curve). }
\label{fig:Lasym}
\end{figure}

{\it(i)} Throughout the analysis we use \mbox{$\mu= 4.3$ GeV} and  $\alpha_s(\mu) = 0.244$,  as was determined
in our numerical study of the  SDE system discussed  in Sec.~\ref{sec:ngene}.

{\it(ii)} For both $\Delta(q^2)$ and  $F(q^2)$,  renormalized at the aforementioned $\mu$,
we employ the fits given by \2eqs{gluonfit}{ghostfit},  respectively.

{\it(iii)} For $\Ls(q^2)$  we employ an excellent fit to the lattice data of~\cite{Aguilar:2021lke}.  The curve is shown in Fig.~\ref{fig:Lasym},  and its functional form is given by
\begin{equation}
\label{Lasymfit}
\Ls(q^2)=F(q^2)T(q^2)+\nu_1 \left( \frac{1}{1+(q^2/\nu_2)^2} - \frac{1}{1+(\mu^2/\nu_2)^2} \right),     
\end{equation}
with
\begin{equation}
    T(q^2) = 1 +\frac{3\lambda_{\srm S}}{4\pi} \left( 1+ \frac{\tau_1}{q^2+\tau_2} \right) \left[ 2\ln \left(  \frac{q^2+ \eta^2(q^2)}{\mu^2+ \eta^2(\mu^2)}\right) + \frac{1}{6}\ln \left( \frac{q^2}{\mu^2} \right) \right],
\end{equation}
and 
\begin{equation}
\label{eta}
    \eta^2(q^2) = \frac{\eta_1^2}{1 + q^2/\eta_2^2}\,,
\end{equation}
where the fitting parameters are given by
\mbox{$\lambda_{\srm S}=0.27$},
 \mbox{$\nu_1=0.179$},  \mbox{$\nu_2 =0.830$ GeV$^{2}$},
 \mbox{$\tau_1= 2.67$ GeV$^{2}$},
 \mbox{$\tau_2  = 1.05$  GeV$^{2}$},
 \mbox{$\eta_1^2= 3.10$   GeV$^{2}$},  and 
 \mbox{$\eta_2^2  = 0.729$  GeV$^{2}$}.

Note that the above fit incorporates,  by construction, the renormalization condition \mbox{$\Ls(\mu^2) = 1$},  corresponding to the soft gluon MOM scheme employed in the lattice simulation of~\cite{Aguilar:2021lke}.
In addition, the zero crossing of $\Ls(q^2)$  is located at about 170 MeV.   

{\it(iv)} 
The value of $\tilde{z}_3$  is determined  from the basic relation given by \1eq{z3from_alphas},
which yields the numerical value \mbox{$\tilde{z}_3 \approx 0.95$}, quoted in \1eq{z3num}.

{\it(v)}  For the form factors $B_1(\ell^2,r^2, \varphi)$ and $B_1(k^2,r^2,\pi-\theta)$ we
interpolate the results for  $B_1(r^2,p^2,\theta_1)$ obtained in Sec.~\ref{sec:ngene} [see  Fig.~\ref{fig:rescoupled}].

\begin{figure}[t]
\begin{minipage}[b]{0.45\linewidth}
\centering
\hspace{-1.0cm}
\includegraphics[scale=0.26]{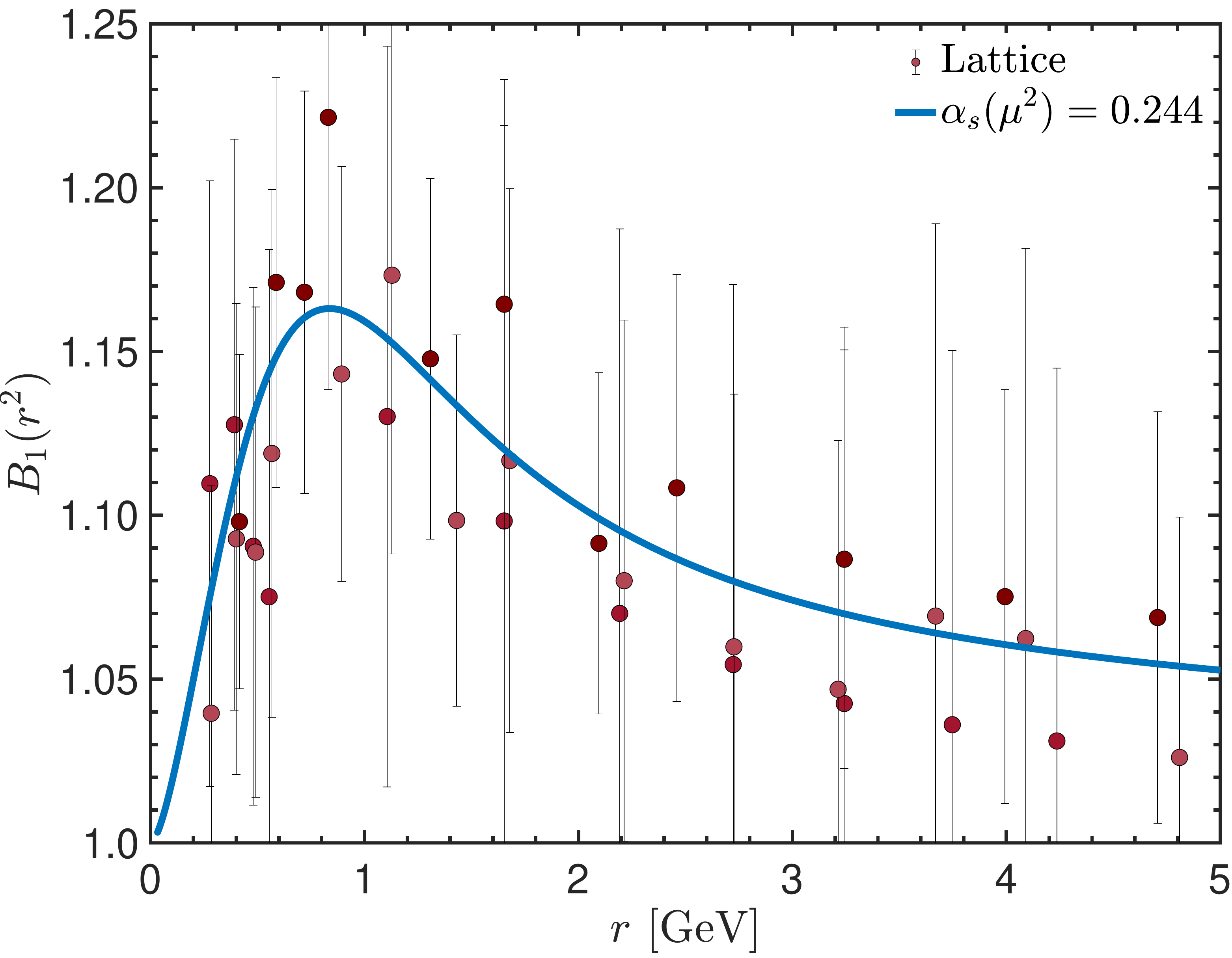}
\end{minipage}
\hspace{0.25cm}
\begin{minipage}[b]{0.45\linewidth}
\includegraphics[scale=0.26]{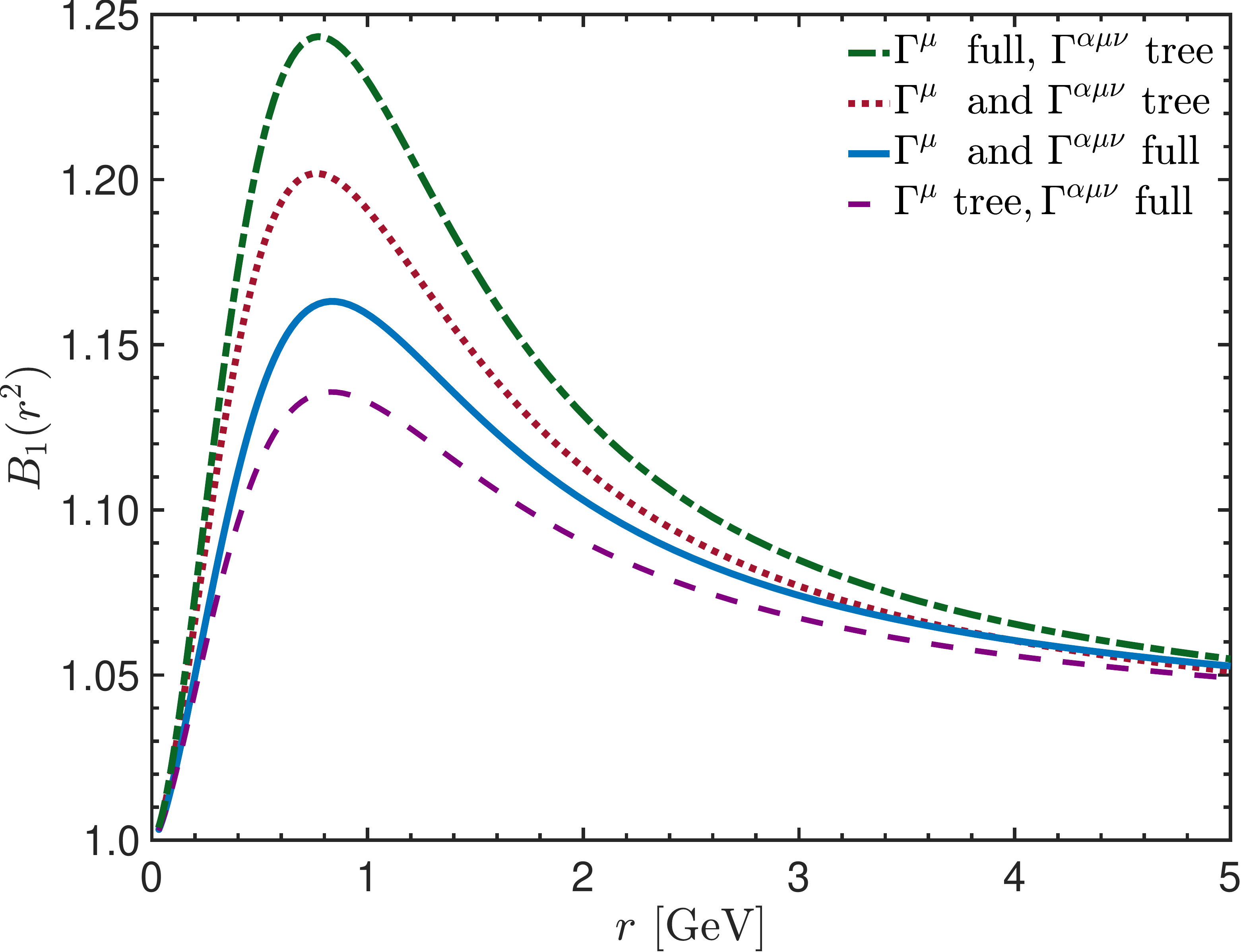}
\end{minipage}
\caption{ Left panel:  The $B_1(r^2)$  obtained as solution  of Eq.~\eqref{FinalEq} (blue continuous)
 together with the lattice data (circles) from~\cite{Ilgenfritz:2006he,Sternbeck:2006rd}.  Right panel:   The numerical impact of dressing the vertices   $\Ls(q^2)$  and  $B_1(\ell^2,r^2, \varphi)$[$B_1(k^2,r^2,\pi-\theta)$]  on $B_1(r^2)$, determined from Eq.~\eqref{FinalEq}. } 
\label{fig:B1soft}
\end{figure}

Using the inputs  described above, 
the $B_1(r^2)$ that emerges as a solution of \1eq{FinalEq} is given by 
the blue continuous curve in the left panel of Fig.~\ref{fig:B1soft}, where it is compared 
 with the $\rm SU(3)$ lattice data of~\cite{Ilgenfritz:2006he,Sternbeck:2006rd}.  
 Although the error bars are rather sizable,  we clearly see that our solution follows the general  trend of the data.  In particular,  notice that both peaks occur in the same intermediate region of momenta.  

The $B_1(r^2)$ may be accurately fitted with the functional form 
\begin{equation} 
\label{fitB1}
    B_1(r^2) = 1 + \frac{r^2(a+br^2)}{1+cr ^2 +dr^4\ln\left[(r^2+ r_0^2)/\rho^2\right]}\,,
\end{equation}
where the parameters are given by \mbox{$a=2.07\, \mbox{GeV}^{-2}$}, $b=9.85\, \mbox{GeV}^{-4}$,  $c=22.3\, \mbox{GeV}^{-2}$,  \mbox{$d=56.4\,  \mbox{GeV}^{-4}$},  \mbox{$r_0^2=1.48\, \mbox{GeV}^2$},  and  \mbox{$\rho^2=1.0\, \mbox{GeV}^2$},
and the  \mbox{$\chi^2/\text{d.o.f.} = 1.0 \times 10^{-6}$}.

We next study the impact that the
amount of ``dressing'' carried by the various vertices  has on $B_1(r^2)$.
To  that end,  we solve \1eq{FinalEq}  considering the three-gluon and ghost-gluon vertices to 
be either at their tree-level values or fully dressed. The results of the four cases considered are displayed 
in the right panel of Fig.~\ref{fig:B1soft}. The hierarchy observed in this plot is
completely consistent with the known infrared properties of these two fundamental vertices: 
at low momenta, the ghost-gluon vertex displays a mild enhancement  
with respect to its tree-level value, whereas the three-gluon vertex is considerably suppressed.  

Based on this particular combination of facts, one would expect that the solution with the
maximal support will be obtained from \1eq{FinalEq} 
when the ghost-gluon vertices are dressed while the three-gluon vertex is kept bare ($ \Ls=1$);
this is indeed what happens, as may be seen from the dot-dashed green curve, which displays
the most pronounced peak. By the same logic, the reverse combination, namely bare ghost-gluon vertices  
and dressed three-gluon vertex, should furnish the most suppressed  $B_1(r^2)$; evidently, this
is what we find, as shown by the purple dashed curve. The remaining cases, where both vertices
are bare, or fully dressed, must lie between the two prior cases; this expectation
is clearly realized within the detailed numerical analysis, as can be seen by the corresponding curves,
indicated by dotted red and continuous blue, respectively.

\begin{figure}[t]
\centering
\includegraphics[scale=0.26]{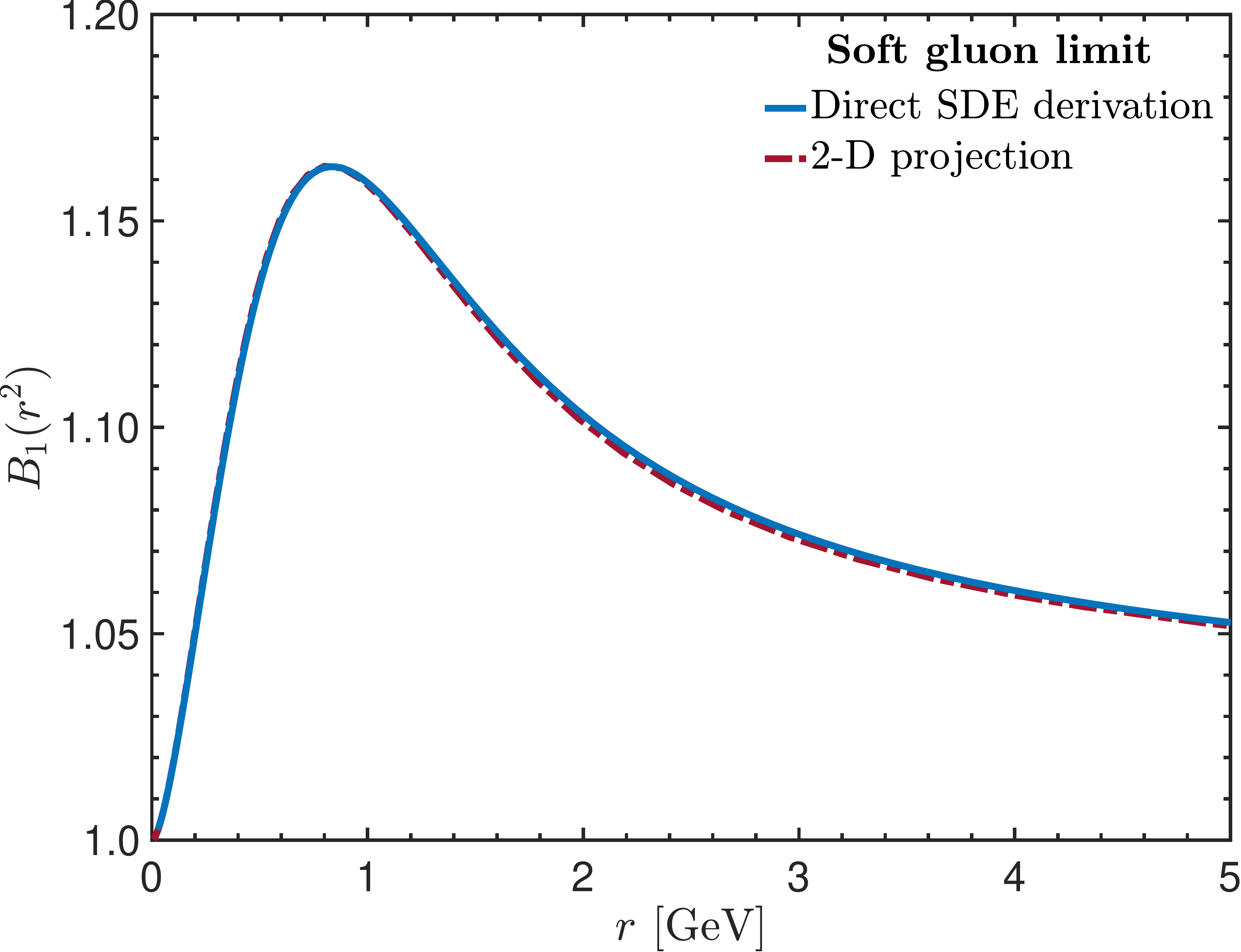}
\caption{Comparison of the soft gluon result obtained as  solution of Eq.~\eqref{FinalEq} (blue continuous curve)  with the one 
extracted from the 3-D plot shown in  Fig.~\ref{fig:conf} (red dashed curve). }
\label{fig:3Dcomp2D}
\end{figure}

We conclude our numerical analysis with an instructive  self-consistency check.
Specifically, as explained in item {\it(v)} above, in order to solve \1eq{FinalEq}
we have used as external input the result for the ghost-gluon vertex for general kinematics, 
derived in Sec.~\ref{sec:ngene}. 
But, as is clear from Fig.~\ref{fig:conf}, the input used
to obtain the soft gluon limit contains already a prediction of what that limit should be,
namely the red dot-dashed curve of $B_1(r^2,q^2,2\pi/3)$,  shown in Fig.~\ref{fig:conf}.
Therefore, a reasonable indication of the self-consistency of the entire procedure
would be the degree of coincidence between the latter  2-D projection  and the result
for $B_1(r^2)$ obtained from \1eq{FinalEq},  namely the blue continuous
curve in either panel of Fig.~\ref{fig:B1soft}.

The direct comparison between these two curves is shown in Fig.~\ref{fig:3Dcomp2D}, where
an excellent coincidence may be observed. Specifically, the maximum discrepancy,
located at about 2 GeV, is of the order of 2\%. The proximity between these results
suggests an underlying consistency between the various ingredients entering in the
corresponding calculations. Note, in particular, that the insertion of lattice ingredients, such as
the gluon propagator and the $\Ls(q^2)$, into the SDEs appears to be a completely congruous operation.

Finally, it is rather instructive to compare  the effective strengths associated with the ghost-gluon and the three-gluon interactions in the soft gluon configuration by means of {\it renormalization-group invariant}  quantities.
To that end, we consider the corresponding effective couplings,
to be denoted by ${\alpha}_{\rm{cg}}(q^2)$ and ${\alpha}_{\rm{3g}}(q^2)$,
defined as (see, \eg ~\cite{Athenodorou:2016oyh,Mitter:2014wpa,Fu:2019hdw})
\be
   {\alpha}_{\rm{cg}}(q^2) = {\alpha}_s(\mu^2) B_1^2(q^2) F^2(q^2)\Dr(q^2)\,,\qquad {\alpha}_{\rm{3g}}(q^2)={\alpha}_s(\mu^2) [\TLs(q^2)]^2 \Dr^3(q^2)\,,  
\label{coup_cg}
\ee
where $\Dr(q^2)$ is the dressing of the gluon propagator, defined in \1eq{gluon},
while $\TLs(q^2)$ is the lattice result of~\cite{Aguilar:2021lke} adjusted to the Taylor scheme, according to 
\2eqs{asytaylor}{z3num}. Note that, by means of this latter adjustment, all ingredients entering in the definitions of both effective couplings
are computed in the same renormalization scheme, namely the Taylor scheme. In addition, 
according to our SDE estimate (see Sec. \ref{sec:ngene}), we have that ${\alpha}_s(\mu) = 0.244$ , at $\mu = 4.3$ GeV.

\begin{figure}[t]
\centering
\includegraphics[scale=0.26]{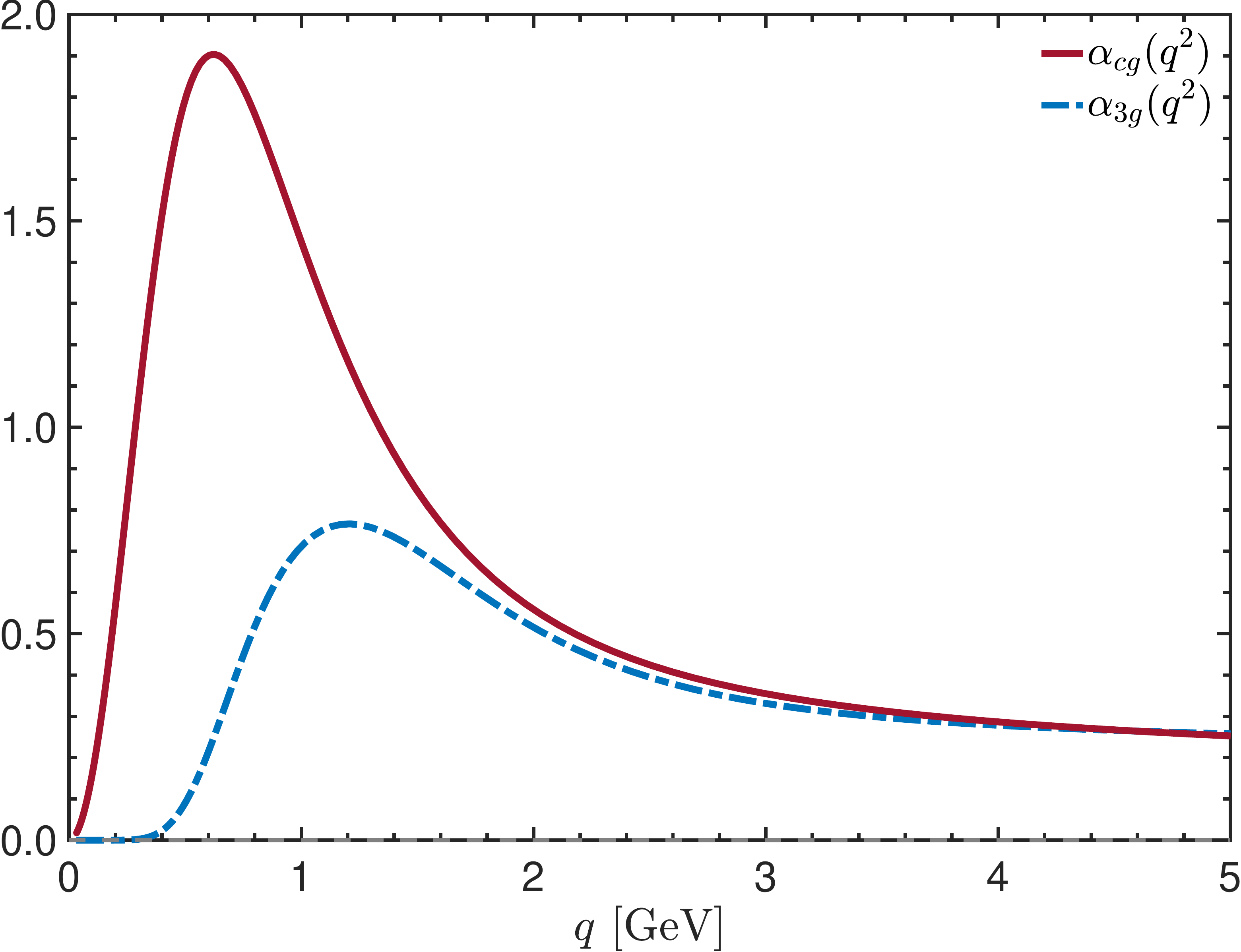}
\caption{ The comparison of the effective couplings, ${\alpha}_{\rm{cg}}(q^2)$ (red continuous line) and ${\alpha}_{\rm{3g}}(q^2)$   (blue dashed).}
\label{fig:coupling}
\end{figure}

The result of the evaluation of the two effective couplings is shown in Fig.~\ref{fig:coupling}. The main feature, consistent with a variety of previous studies~\cite{Huber:2012kd,Blum:2014gna,Williams:2014iea,Cyrol:2016tym,Cyrol:2017ewj, Aguilar:2020yni},
is the considerable suppression displayed by ${\alpha}_{\rm{3g}}(q^2)$ in the region below 2 GeV.


\section{\label{sec:conc}Conclusions}

In this work we have carried out a thorough study of the dynamics related with the
ghost sector of pure Yang-Mills theories, incorporating into the
standard SDEs pivotal elements stemming from recent lattice studies.
In fact, these lattice results serve both as external inputs for some of the quantities that
are difficult to determine accurately within the SDE framework, such as the gluon propagator and certain
components of the three-gluon vertex,  as well as refined benchmarks for testing the reliability of our numerical solutions,  such as the ghost dressing function. 
Specifically, the lattice gluon propagator has been used as a global input in all SDEs considered in the present study, while, in the ``soft gluon'' SDE, the lattice data for the three-gluon vertex 
in the same limit have been employed.

The main results of our analysis are 
succinctly captured in  Figs.~\ref{fig:rescoupled} and~\ref{fig:3Dcomp2D}. 
In particular,  in the left panel of Fig.~\ref{fig:rescoupled},
the ghost dressing function obtained as a solution of the coupled SDE system is compared to the 
results of the lattice simulation of~\cite{Boucaud:2018xup}. 
It is important to appreciate that the success of this comparison 
hinges on the optimization for the cure of discretization artifacts, in connection to the scale-setting and continuum extrapolation, imposed on this set of lattice data.
Indeed, the difference between the latter lattice data and the (non-extrapolated)  
results of~\cite{Bogolubsky:2009dc}, displayed in Fig.~\ref{fig:ghostdress},
is rather substantial, affecting a phenomenologically important region of momenta. 
This difference accounts almost entirely for the discrepancies
found in earlier studies~\cite{Aguilar:2009pp,Aguilar:2013xqa,Aguilar:2018csq},  where the SDE results were compared with the data of~\cite{Bogolubsky:2009dc}.  

We next turn to Fig.~\ref{fig:3Dcomp2D}, where the curves for $B_1(r^2)$,  
obtained following two distinct procedures, are compared.   
The excellent agreement between both results suggests an underlying self-consistency
among the several elements participating non-trivially in the computation of these results.
Particularly interesting in this context is the pivotal role played by the three-gluon vertex,
which appears in both computations leading to the results of Fig.~\ref{fig:3Dcomp2D},
albeit in rather dissimilar kinematic arrangements. 
Specifically, to obtain the result marked by the blue continuous
curve, the vertex was approximated by its classical tensor structure,
accompanied by the corresponding form factors in general kinematics, as explained in Sec.~\ref{sec:coupled}.
Instead, the red dashed curve is obtained through the direct use of the
lattice results in the soft gluon limit, according to the discussion in Sec.~\ref{sec:cgsoft}. The coincidence between
the results indicates that the STI-based construction of~\cite{Aguilar:2019jsj},
which gave rise to the form factors used for the
computation of the blue continuous curve, is quite reliable.
In that sense, it is rather gratifying to see how well the
dynamical equations respond in this particular set of circumstances; in fact,
the use of lattice data as SDE inputs appears to be completely consistent. 

Note that the present study is fully compatible with the assertion of~\cite{Huber:2017txg,Huber:2018ned}
that the four-point function represented by the yellow ellipse in Fig.~\ref{fig:diagvert} 
is numerically rather negligible.
Evidently, the excellent agreement with the lattice
found in the left panel of Fig.~\ref{fig:rescoupled} indicates that the omission of the corresponding term 
from the skeleton expansion of the SDE kernel does not introduce any appreciable error. 
In fact, it is interesting to observe that an entirely different conclusion about the importance of this
four-point function would have been drawn 
if the non-extrapolated lattice results of~\cite{Bogolubsky:2009dc}  had been used for the
comparison in Fig.~\ref{fig:rescoupled}.
Specifically, any attempt to interpret the difference alternatively as a consequence of the kernel truncation would force this four-point function to acquire considerably higher values than those found in the detailed analysis of~\cite{Huber:2017txg,Huber:2018ned}.

Finally, it would be interesting to extend the considerations of Sec.~\ref{sec:cgsoft} to the case of the
quark-gluon vertex, whose SDE and corresponding skeleton expansion
are given by replacing in Figs.~\ref{fig:coupsys} and~\ref{fig:diagvert}, respectively, all ghost lines by quark lines. 
In particular, the soft gluon limit of the quark-gluon vertex
involves three form factors, whose determination 
has attracted particular attention over the years.  In fact, up until very recently~\cite{Kizilersu:2021jen},
notable discrepancies existed between the continuous predictions~\cite{Bhagwat:2004kj,LlanesEstrada:2004jz,Fischer:2003rp,Fischer:2006ub,Aguilar:2014lha,Aguilar:2016lbe,Oliveira:2018fkj, Oliveira:2018ukh,Oliveira:2020yac} and the results of lattice simulations~\cite{Skullerud:2002ge,Skullerud:2003qu,Skullerud:2004gp,Lin:2005zd,Kizilersu:2006et,Oliveira:2016muq,Sternbeck:2017ntv}.
It is likely that the inclusion of $\Ls$ in the SDE treatment might shed further light on this intricate problem.

\section*{\label{sec:acknowledgments}Acknowledgments}
The work of  A.~C.~A. is supported by the CNPq grant 307854/2019-1 and the project 464898/2014-5 (INCT-FNA).
A.~C.~A. ,  C. ~O.~A,  and M.~N.~F.    also acknowledge financial support from the FAPESP projects 2017/05685-2,  2019/05656-8, and 2020/12795-1, respectively.
J.~P. is supported by the  Spanish MICIU grant FPA2017-84543-P,
and the  grant  Prometeo/2019/087 of the Generalitat Valenciana. 
F.~D.~S. and J.~R.~Q.~ are supported the Spanish MICINN grant PID2019-107844-GB-C2, and regional Andalusian project P18-FR-5057.   This study was financed in part by the Coordena\c{c}\~{a}o de Aperfei\c{c}oamento de Pessoal de N{\'\i}vel Superior - Brasil (CAPES) Finance Code 001 (B.~M.~O).

\newpage 

\appendix

\section{\label{sec:App_renor} Taylor and soft gluon renormalization schemes}

In this Appendix we review certain basic relations that are necessary for the meaningful comparison of results
obtained within two special renormalization schemes, namely the Taylor scheme~\cite{Boucaud:2011eh} and the soft gluon scheme~\cite{Aguilar:2021lke}.

The general relations connecting bare and renormalized quantities are given by 
\begin{align}
\Delta_{\srm R}(q^2) & = Z_{\rm A}^{-1}\Delta(q^2)\,,  \qquad \quad \Gamma_{\!\srm R}^\nu(r,p,q)  = Z_{1}\Gamma^\nu(r,p,q)  \,, \ \nonumber \\
F_{\!\srm R}(q^2) & = Z_{\rm c}^{-1} F(q^2)\,,   \qquad  \quad 
\Gamma^{\alpha\mu \nu}_{\!\srm R} (q,r,p)  =  Z_{3} \,\Gamma^{\alpha\mu \nu}(q,r,p) \,, \nonumber \\
g_{\srm R}  &= Z_{g}^{-1} g \,,\,\,\,\,\, \qquad\qquad Z_{ g} ^{-1}  = Z_{1}^{-1} Z_{\rm A}^{1/2} Z_{\rm c}  =  Z_{3}^{-1} Z_{\rm A}^{3/2}\,,
\label{renorm}
\end{align}
where  $Z_{\rm A}$,  $Z_{\rm c}$, $Z_{1}$, $Z_{3}$,  and $Z_{g}$ are the  corresponding renormalization constants.
In what follows we will reserve this notation for the renormalization constants in the Taylor scheme, while 
the corresponding constants in the soft gluon scheme will carry a ``tilde''.   

Clearly, both schemes impose on propagators the typical MOM condition, \ie 
\begin{equation}
  \Delta_{\srm R}^{-1}(\mu^2) = {\mu^2} \,, \qquad \qquad F_{\!\srm R}(\mu^2) = 1 \,.
\label{propsMOM}
\end{equation}

The difference between the two schemes manifests itself at the level of the renormalization conditions applied
on vertex form factors. The Taylor scheme is motivated by the corresponding theorem~\cite{Taylor:1971ff}, which states that the bare ghost-gluon vertex reduces to tree level in the soft ghost limit, $p = 0$, \ie $\Gamma^\nu(r,0,-r) = r^\nu$. Since the bare $\Gamma^\nu(r,0,-r)$ is finite, so is the corresponding renormalization constant, $Z_{1}$.  Then, the Taylor scheme is defined by imposing  that the renormalized vertex also reduces to tree level at $p = 0$, \ie $\Gamma_{\!\srm R}^\nu(r,0,-r) = r^\nu$, implying through \1eq{renorm} that $Z_{1} = 1$. At the level of the tensor decomposition
introduced in \1eq{decomp}, this implies 
\begin{equation}
\Gamma_{\!\srm R}^\nu(r,0,-r) = \left[ B_1(r,0,-r) - B_2(r,0,-r) \right] r^\nu \,,
\end{equation}
from which follows that, in the Taylor scheme, $B_1(r,0,-r) - B_2(r,0,-r) = 1$. Note that this scheme does not impose any condition on the \emph{individual} $B_1$ and $B_2$; in particular, $B_1(\mu^2) \neq 1$, as may be
clearly appreciated in Fig.~\ref{fig:B1soft}.
Instead, by definition of the soft gluon scheme, the $\Ls(q^2)$   
determined on the lattice satisfies \mbox{$\Ls(\mu^2) = 1$}~\cite{Aguilar:2021lke}.

Thus, in order to self-consistently incorporate the lattice results of~\cite{Aguilar:2021lke} 
into the computation of $B_1(r^2)$ through \1eq{FinalEq}, we must 
relate the  $\Ls(q^2)$ determined in the soft gluon scheme with the corresponding quantity 
in the Taylor scheme,  to be denoted by $\TLs(q^2)$.

In general, the transition between two renormalization schemes is implemented by means of finite
renormalization constants, which relate both the renormalized quantities as well as the renormalization
constants. In particular, $\Ls(q^2)$ and  $\TLs(q^2)$, as well as 
the renormalization constants $Z_3$ and ${\widetilde Z}_3$,
are related by a finite renormalization constant, $\tilde{z}_3$, according to~\cite{Aguilar:2020yni}
\be
\Ls(q^2) =  \tilde{z}_3\TLs(q^2)\,, \qquad  \tilde{z}_3 = \frac{{\widetilde Z}_3}{Z_3}\,.
\label{asytaylor} 
\ee

Given that the unrenormalized (bare) gauge coupling is identical in both schemes,
and that $Z_A = {\widetilde Z}_A$, the relations on the last line of \1eq{renorm} yield 
\be
\frac{g_{\srm R}}{\tilde{g}_\srm{R}} = \frac{\tilde{Z}_g}{Z_g} = \frac{{\widetilde Z}_3}{Z_3} = \tilde{z}_3\,,
\label{ratios}
\ee
or,  equivalently,  in terms of the corresponding charges one has 
\be 
\tilde{z}_3 = \sqrt{\frac{{\alpha}_s(\mu)}{{\widetilde \alpha}_s(\mu)}}.
\label{z3from_alphas}
\ee

\1eq{z3from_alphas} permits a nonperturbative estimate of $\tilde{z}_3$, given that reliable information 
on the values of both $\alpha_s(\mu)$ and $\widetilde{\alpha}_s(\mu)$, at $\mu =4.3$ GeV, is available.
Specifically, the value of $\alpha_s(\mu)$ obtained from the analysis of Sec.~\ref{sec:coupled}
is $\alpha_s(\mu)=0.244$, while in the lattice simulation that produced the $\Ls(q^2)$ the value
of the charge was determined to be ${\widetilde \alpha}_s(\mu) = 0.27$. 
Consequently, from \1eq{z3from_alphas} we obtain
\be 
\tilde{z}_3 \approx 0.95.
\label{z3num}
\ee

\section{\label{sec:App_latt} Lattice artifacts and scale-setting} 

In this Appendix we discuss in detail the method employed for curing the lattice data
of the gluon and ghost propagator from lattice artifacts, and elaborate on the procedure
leading to the appropriate setting of the scale.  

One of the relevant improvements in our current analysis stems from the high quality of the quenched data we used here for the gluon propagator. This relies on the following three implementations: {\it(i)} a careful and efficacious treatment removing discretization artifacts from our data (continuum extrapolation)~\cite{Boucaud:2018xup}; {\it(ii)} a cure from finite-size effects (infinite volume extrapolation) which capitalizes on the use of seven different volumes in physical units; and {\it(iii)} a sensible combination of our data with those of~\cite{Bogolubsky:2009dc} for a better control of the systematic effects and increase of statistics.

{\it(i)} The issue of the continuum extrapolation is handled as in~\cite{Boucaud:2018xup}. There, through an analysis of lattice simulations with five different bare couplings ($\beta=6/g^2(a)$=5.6, 5.7, 5.8, 5.9, and 6.0), it was demonstrated that, for any two of them (say, $\beta$ and $\beta_0$), 
\begin{equation}\label{eq:CGamma}
\frac{\Dr_{\s L}(q^2,\mu^2;a(\beta))}{\Dr_{\s L}(q^2,\mu^2;a(\beta_0))} =  1 + 
a^2(\beta) \left[ 1 - \frac{a^2(\beta_0)}{a^2(\beta)} \right]  \left[ c\,( q^2 - \mu^2)  + d\, \left( \frac{q^{[4]}}{q^2}  -\frac{\mu^{[4]}}{\mu^2} \right) \right] \; + \dots \,,
\end{equation}
where $\Dr_{\s L}$ represents the MOM-scheme gluon dressing function at fixed cutoff, 
\begin{equation}
\label{eq:GammaL}
\Dr_{\s L}(q^2,\mu^2;a) = \frac{\Dr(q^2,a)}{\Dr(\mu^2,a)} = \Dr_{\rm{\s R}}(q^2,\mu^2) + {\cal O}(a^2) \,.       
\end{equation}
$\Dr(q^2,a)= q^2 \Delta(q^2,a)$ is the bare dressing introduced in \1eq{gluon}, $q$ represents the gluon momentum, $\mu$ is the subtraction point, and $a$ denotes the lattice spacing. $\Dr_{\s R}$ in \1eq{eq:GammaL} stands for the renormalized dressing function after removing the cutoff by extrapolation to the continuum limit, $a \to 0$. Note that the subtraction procedure alone, without taking this limit, cannot prevent the answer from exhibiting a residual dependence on $a$,  captured by the ${\cal O}(a^2)$ in \1eq{eq:GammaL}. 
In addition, $q^{[4]}:=\sum\limits_{i=1}^{4} q_i^4$ (the same for $\mu$) stands for the first $H(4)$ invariant of the extrapolation that cures the hypercubic artifacts~\cite{Becirevic:1999uc,Becirevic:1999hj,deSoto:2007ht,Catumba:2021hcx}; $c$ and $d$ are dimensionless coefficients, and the ellipses indicate corrections of order ${\cal O}(a^4)$, might they be $O(4)$-invariant contributions or those originating from higher $H(4)$ invariants (which can be properly neglected if $a^2 q^2$ is sufficiently small).

It is clear from \1eq{eq:CGamma} that estimates from simulations differing in their discretization deviate from each other by corrections which depend on the differences of their lattice spacings and, hence, can only coincide after continuum extrapolation. This extrapolation is implemented in~\cite{Boucaud:2018xup} as follows: one first determines $c$ and $d$ by fitting the results from all the
simulations involved; next, one takes $a(\beta_0) \to 0$ $(\beta_0 \to \infty)$ in \1eq{eq:CGamma} and obtains from it $\Dr_L(q^2,\mu^2,0)$ for each data from each simulation, identifying the answer as the extrapolated value, $\Dr_{\rm{\s R}}(q^2,\mu^2)$ as given in \1eq{eq:GammaL}. Following this procedure 
for all simulations reported in Table~\ref{tab:setup}, we deliver the gluon propagator data in the physical continuum limit, shown in the left panel
of Fig~\ref{fig:X1}. The data clearly exhibit the expected physical scaling over a wide range of momenta, thus reinforcing the reliability of the entire method.

\begin{table}
\begin{tabular}{|c|c|c|c|c|c|c|c|}
\hline 
\hline 
$\quad\beta\quad$ & $\quad5.6\;$ & $\quad5.6\;$ &$\quad 5.7\;$ &$\quad5.8\;$ & $\quad5.8\;$ & $\quad5.9\;$ & $\quad5.9\;\,$ \\
\hline 
$L$ & 48 & 52 & 40 & 32 & 48 &30& 64  \\ 
\hline
\hline 
\end{tabular} \;. 
\caption{Seven setups employed  in our analysis to deal  with finite volume artifacts (four of them were already exploited in Ref.\cite{Boucaud:2018xup}). $L$ stands for the lattice size in units of the lattice spacing, $a(\beta)$. }
\label{tab:setup}
\end{table}

\begin{figure}[t]
\begin{minipage}[b]{0.45\linewidth}
\centering
\hspace{-1.0cm}
\includegraphics[scale=0.26]{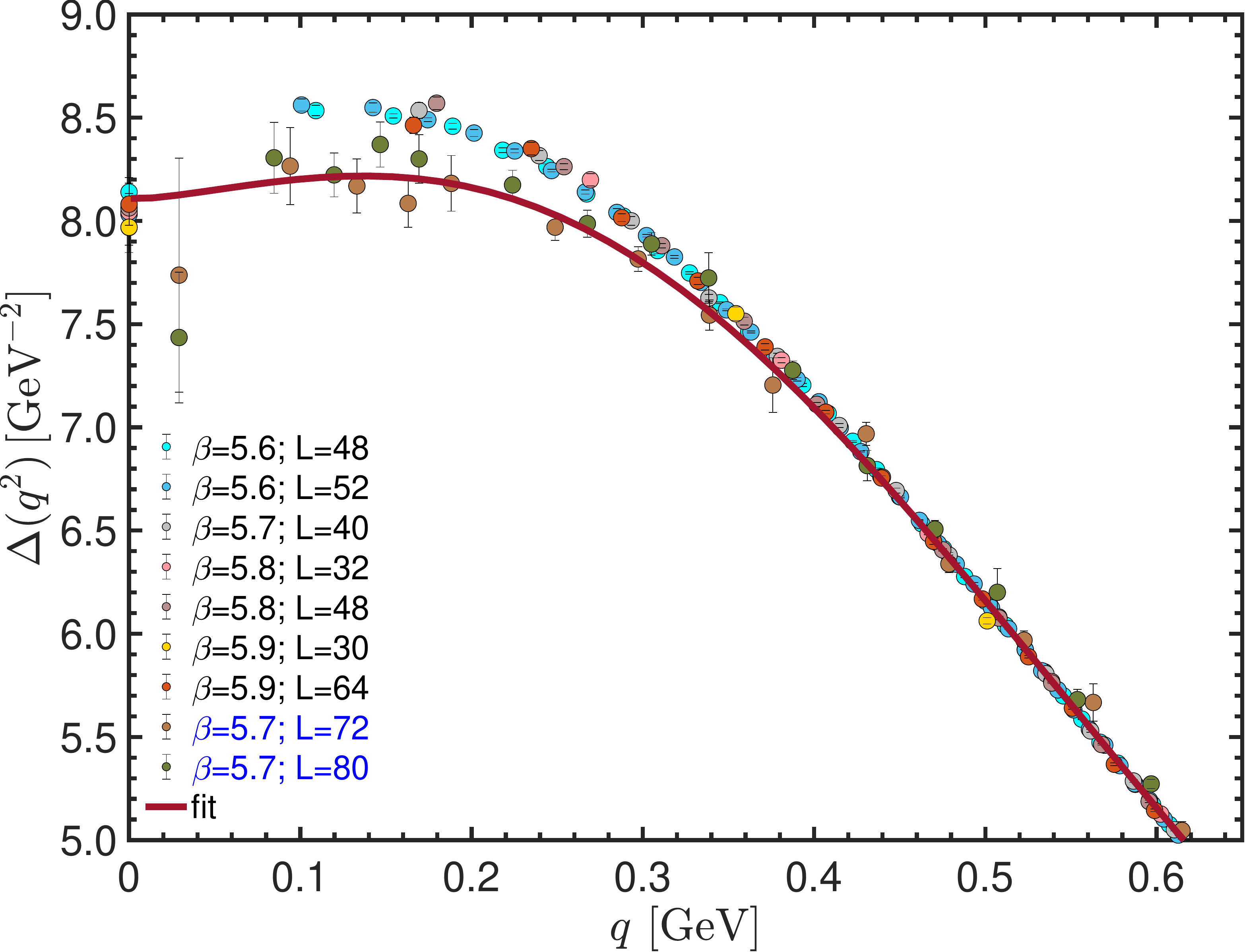}
\end{minipage}
\hspace{0.25cm}
\begin{minipage}[b]{0.45\linewidth}
\includegraphics[scale=0.26]{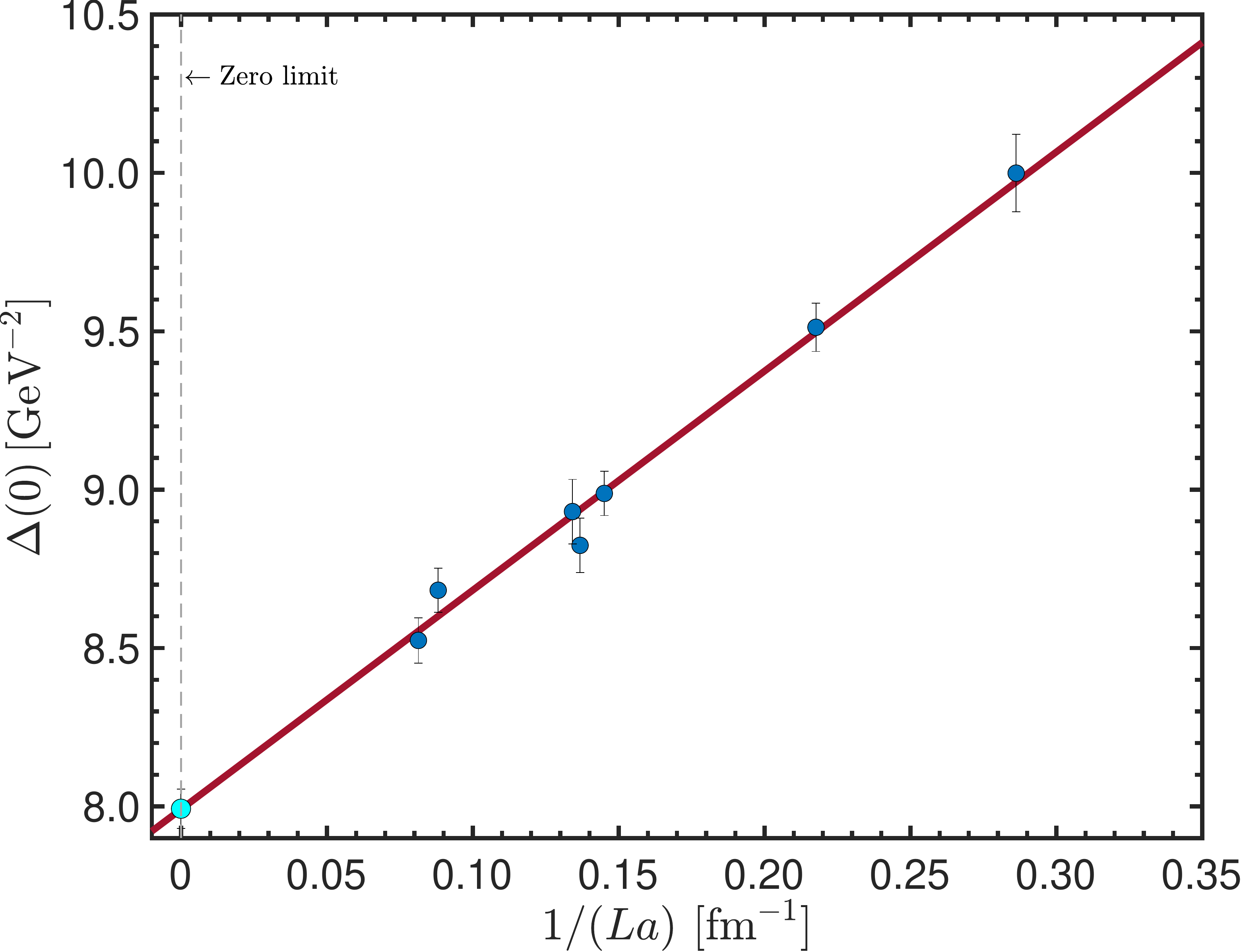}
\end{minipage}
\caption{ Left panel: Lattice data for the gluon propagator,  $\Delta(q^2)$, obtained with the setups reported in Table~\ref{tab:setup}, after continuum and infinite volume extrapolations, and subsequently combined with the data from~\cite{Bogolubsky:2009dc} (blue legend) after implementing on them the scale resetting procedure.  The red continuous line
represents the fit given by  Eq.~\eqref{gluonfit}.  Right panel: Zero-momentum lattice estimates of the gluon propagator obtained from all the lattice simulations quoted in Table~\ref{tab:setup}. The solid line corresponds to 
a linear fit consistent with Eq.\,\eqref{eq:Vinf}, specialized for $q=0$.}   
\label{fig:lattprop}
\end{figure}

{\it(ii)} For the purpose of dealing with finite-size effects, we have capitalized on the seven different volumes (in physical units) obtained with the simulations reported in Table~\ref{tab:setup}. Gluon propagators computed from them, and extrapolated to the physical continuum limit, are seen to differ only at very low momenta, where the data behave as
\begin{equation}\label{eq:Vinf}
\Delta(q^2,L) = \Delta(q^2,\infty) \left[ 1 + \left(\frac{c_1}{L a(\beta)}\right)  \exp \bigg(-c_2 L a(\beta) q \bigg) \right] \,.
\end{equation}
In particular, specializing to zero momentum, Eq.\,\eqref{eq:Vinf} predicts a linear dependence of
the finite volume lattice results on $1/L a(\beta)$, which, as shown in  the right panel of Fig.\,\ref{fig:lattprop}, is nicely displayed by all our data. We can therefrom estimate for the infinite volume zero-momentum gluon propagator: $\Delta(0,\infty)=7.99(5)$ GeV$^{-2}$. 

With this latter result in hand, we have applied Eq.\,\eqref{eq:Vinf} to our continuum-limit, finite volume data sets for all momenta
$q \leq 0.5$ GeV (above this momentum, the impact of finite volume corrections is plainly negligible). We have thus fitted the two parameters, $c_1$=3.6 GeV$^{-1}$, $c_2$=0.27, and computed $ \Delta(q^2,\infty)$ for every lattice estimate of the gluon propagator within the low momenta window. In this way, we have produced the data depicted in the left panel of Fig.\,\ref{fig:lattprop} for $q \leq 0.5$ GeV.

{\it(iii)} Finally, aiming at increased statistics and a better control of the systematics, we would like to supplement our results with those from large-volume simulations ($64^4$, $72^4$, and $80^4$ lattice sites) at $\beta$=5.7, taken from~\cite{Bogolubsky:2009dc}. However, a complication arises: we have used the scale-setting procedure, or ``\emph{calibration}", described in~\cite{Boucaud:2018xup}, which is different from that applied in~\cite{Bogolubsky:2009dc}; in addition, in~\cite{Bogolubsky:2009dc}   
no continuum extrapolation was carried out. This is an important issue, because the calibration, implemented by imposing that a given lattice \emph{observable} acquires its physical value, depends on the choice of the observable. The latter was made abundantly clear in~\cite{Boucaud:2018xup} through the comparison of the ratios $a(\beta)/a(\beta_0)$ obtained by applying the scale-setting methods based on Sommer's parameter (heavy quark potential) and on the Taylor coupling: they differ from each other, but converge when $\beta \to \infty$ (continuum limit). In conclusion, results obtained with different calibrations can coincide 
only after taking the continuum extrapolation, and the deviations between them (before this limit is taken) 
can be thus interpreted as a discretization artifact.  

All the above has been explicitly shown in~\cite{Duarte:2016iko,Boucaud:2017ksi,Duarte:2017wte}, where a simple but effective remedy
for correcting these deviations has been proposed: a scale resetting $\bar{a}(\beta)=(1+\delta) a(\beta)$ is to be applied to the non-extrapolated data such that they match the extrapolated ones,  thus recovering the physical scaling. We therefore define~\cite{Boucaud:2017ksi}
\begin{equation}
\Dr_{\rm{\s R}}(q^2,\mu^2) =\Dr_{\s L}\left( \frac{\bar{a}^2(\beta)}{a^2(\beta)} q^2, \frac{\bar{a}^2(\beta)}{a^2(\beta)} \mu^2;a(\beta)\right) = \frac{\Dr((1+\delta)^2 q^2,a(\beta))}{\Dr((1+\delta)^2\mu^2,a(\beta))} \;,
\end{equation}

\noindent
where a \emph{recalibration} $a(\beta) \to \bar{a}(\beta)$ is performed, such that the scale setting for $\bar{a}(\beta)$ is assumed to rely on the continuum gluon propagator, thereby implying that $c$=$d$=0. In practice, we adjust the parameter $\delta$ such that the \emph{recalibrated} gluon propagator data from~\cite{Bogolubsky:2009dc} and ours optimally agree in the entire range  of available momenta. We thus obtain $\delta$=0.08 and are left with the results displayed in the left panels of Figs.~\ref{fig:X1} and \ref{fig:lattprop}. The agreement is excellent for all momenta roughly above \mbox{0.35 GeV}, while  below it is still acceptable. Importantly, data obtained from applying both approaches exhibit the same key feature: the derivative changes its sign around a maximum located roughly at \mbox{0.15 GeV}.

These considerations allow us to exploit  
all data (including those recalibrated from~\cite{Bogolubsky:2009dc}) and show that they can be fitted rather accurately over the entire range of momenta (see left panel of Fig.\,\ref{fig:X1}) by the following functional form\footnote{When data differ slightly at very low momenta, those estimated with smaller volumes have been discarded from the fit.} 
\be 
\Delta^{-1}(q^2) =  q^2 \left[ 1 + \left( \kappa_{1}  - \frac{\kappa_{2} }{ 1 + (q^2/\kappa_{4}^2)^2}  \right) \ln\left( \frac{q^2}{\mu^2} \right) \right]+ R(q^2) - R(\mu^2)  \,,
\label{gluonfit}
\ee
with 
\be 
R(q^2) =\frac{ \sigma_{0} + \sigma_{1} q^2 }{1 + (q^2/ \sigma_{2}^2 )  +  (q^2/\sigma_{4}^2)^2 } \,, 
\ee
where the fitting parameters are  
\mbox{$ \kappa_{1} = 0.114$},   
\mbox{$ \kappa_{2} = 0.0252$},    
\mbox{$\kappa_{4}^2 = 4.926\, $GeV$^{2}$}, 
\mbox{$\sigma_{0} = -0.406\, $GeV$^{2}$},  
\mbox{$ \sigma_{1} = -0.518$},   
 \mbox{$ \sigma_{2}^2=  10.266\,$GeV$^{2}$},  and,    
 \mbox{$\sigma_{4}^2 = 4.631\, $GeV$^{2}$}.

\begin{figure}[t]
\begin{minipage}[b]{0.45\linewidth}
\centering
\hspace{-1.0cm}
\includegraphics[scale=0.26]{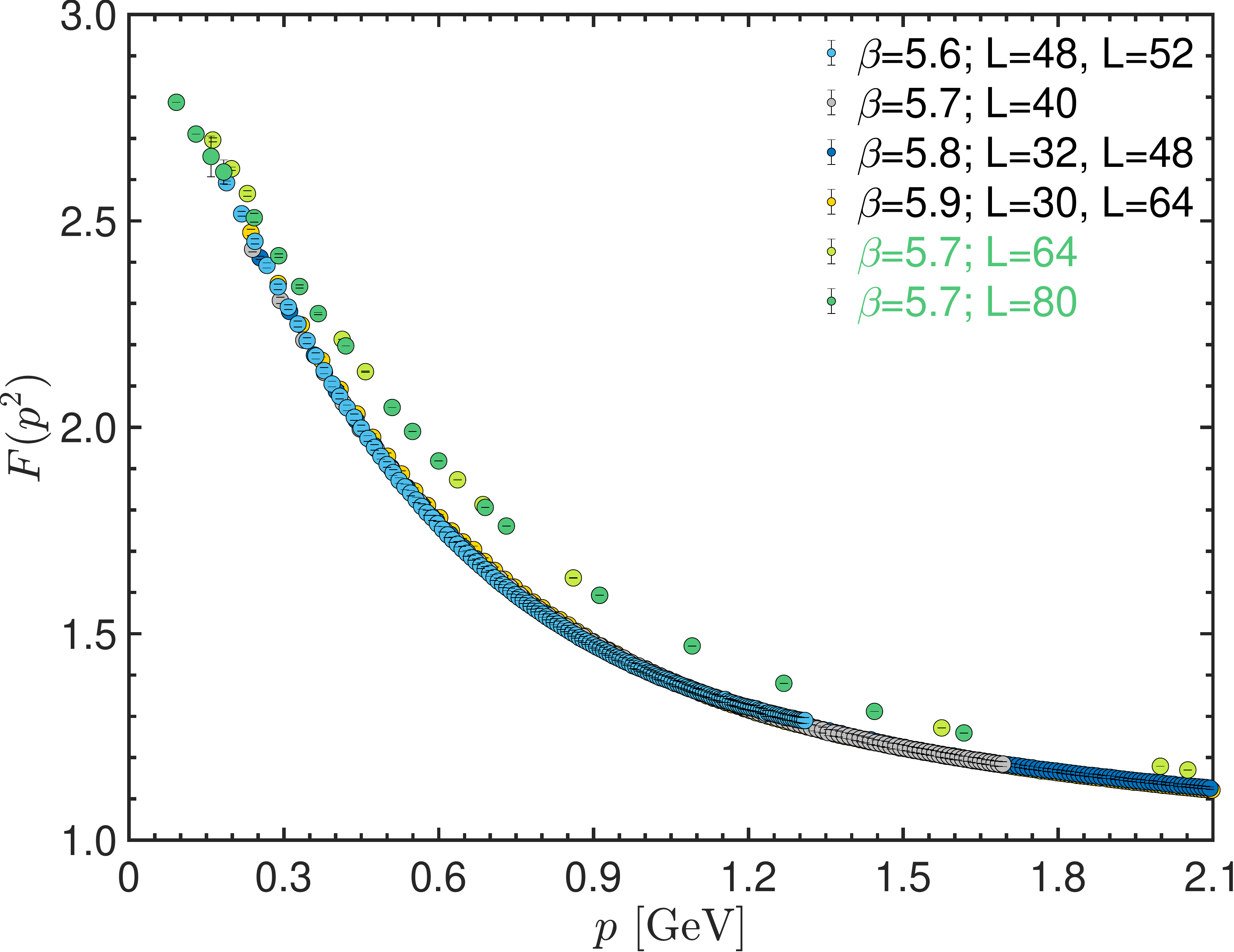}
\end{minipage}
\caption{ The lattice data for the ghost dressing function from~\cite{Boucaud:2018xup}(blue points and legend) compared 
with the corresponding data of~\cite{Bogolubsky:2009dc} (green points and black legend).} 
\label{fig:ghostdress}
\end{figure}

For the ghost dressing function, which is not an input but rather a benchmark for the numerical solution of the corresponding SDE, we have considered the data of~\cite{Boucaud:2018xup}, which have undergone the same extrapolation to the physical continuum limit explained above.
Their comparison with the data from~\cite{Bogolubsky:2009dc} (without scale resetting), shown by the 
green points in the Fig.~\ref{fig:ghostdress},
makes very apparent the importance of the continuum limit. The solution of the ghost SDE, while it misses the data at fixed cutoff, reproduces very well the behavior of extrapolated ones, as shown in Fig.\,\ref{fig:rescoupled}. The resulting dressing function can be accurately fitted by the following functional form
\begin{equation}
    F^{-1}(p^2) = 1 +\frac{9 \lambda_{\srm F}}{16\pi} \left( 1+ \frac{\rho_1}{p^2+\rho_2} \right) 
    \ln\left( \frac{p^2+ \eta^2(p^2)}{\mu^2+ \eta^2(\mu^2)}\right), 
\label{ghostfit}    
\end{equation}
with  $\eta^2(q^2)$ given by  Eq.~\eqref{eta} and  the fitting parameters fixed  at the values   \mbox{$\lambda_{\srm F}=0.22$}, 
 \mbox{$\rho_1= 6.34$ GeV$^{2}$},
 \mbox{$\rho_2  = 2.85$  GeV$^{2}$},
 \mbox{$\eta_1= 0.107$   GeV$^{4}$,} and 
 \mbox{$\eta_2  = 11.2$  GeV$^{2}$}. 



%

\end{document}